\documentclass[10pt,twocolumn]{article}
\usepackage[top=2cm, bottom=2cm, left=1.85cm, right=1.85cm]{geometry}
\usepackage{times}  %
\usepackage[sort&compress,numbers]{natbib}  %
\usepackage[hyphens]{url}
\usepackage{graphicx}  %
\usepackage[keeplastbox]{flushend}
\usepackage[hyphens]{url}  %
\usepackage{graphicx} %
\usepackage{tikzsymbols}

\newcommand{\descr}[1]{\smallskip\noindent\textbf{#1}}

\usepackage{sectsty}
\sectionfont{\bfseries\Large\raggedright}

\usepackage{url}                                %
\usepackage{array,multirow,graphicx,adjustbox}  %
\usepackage{booktabs}                           %
\usepackage[utf8]{inputenc}
\usepackage[ruled,algosection,noend,linesnumbered]{algorithm2e}
\usepackage{float}
\usepackage{paralist}
\usepackage{amsmath}
\usepackage{amsfonts}
\usepackage{xcolor}
\usepackage{csquotes} 
\usepackage{tabularx}
\usepackage{makecell}
\usepackage{pbox}
\usepackage[hang,flushmargin]{footmisc}
\usepackage{flushend}
\usepackage{xspace}
\usepackage{multicol, lipsum}
\usepackage[T1]{fontenc}

\usepackage{xcolor}
\usepackage{booktabs}
\usepackage{graphicx}
\usepackage{paralist}
\usepackage[small,bf]{caption}
\usepackage{subfigure}

\let\oldbibliography\thebibliography
\renewcommand{\thebibliography}[1]{%
  \oldbibliography{#1}%
  \setlength{\itemsep}{2pt}%
}

\usepackage[hang,flushmargin]{footmisc}

\usepackage[compact]{titlesec}
\titlespacing*{\section}{0pt}{*3}{3pt}
\titlespacing*{\subsection}{0pt}{*2}{2pt}
\usepackage{xspace}

\makeatletter
\def\url@leostyle{%
  \@ifundefined{selectfont}{\def\UrlFont{}}%
  {\def\UrlFont{}}%
}
\makeatother
\usepackage[hyphenbreaks]{breakurl}

\usepackage[bookmarks=true, bookmarksnumbered=true, colorlinks=true, linkcolor=linkcol, citecolor=citecol, urlcolor=urlcol, hypertexnames=true]{hyperref}

\definecolor{darkgreen}{RGB}{0, 100, 0}
\definecolor{linkcol}{rgb}{0.3,0,0}
\definecolor{citecol}{rgb}{0.3,0,0}
\definecolor{urlcol}{rgb}{0.3,0,0}

\makeatletter
\def\url@leostyle{%
  \@ifundefined{selectfont}{\def\UrlFont{\small}}%
  {\def\UrlFont{}}%
}
\makeatother
\newif\ifcomment
\commentfalse
\ifcomment
\newcommand{\sz}[1]{{\bf \textcolor{brown}{SZ: #1}}}
\newcommand{\jbnote}[1]{{\bf \textcolor{magenta}{JB: #1}}}
\newcommand{\edc}[1]{{\bf \textcolor{blue}{EDC: #1}}}
\newcommand{\gs}[1]{{\bf \textcolor{red}{GS: #1}}}
\newcommand{\pp}[1]{{\bf \textcolor{orange}{PP: #1}}}
\else
\newcommand{\sz}[1]{}
\newcommand{\jbnote}[1]{}
\newcommand{\edc}[1]{}
\newcommand{\gs}[1]{}
\newcommand{\pp}[1]{}
\fi

\newcommand{\totalusers}{856,222 }
\newcommand{\inactiveusers}{436,949 }
\newcommand{\datasetstart}{July 1, 2021 }
\newcommand{\datasetend}{August 4, 2021}

\newcommand{\percentagelangen}{90.1\% }
\newcommand{\percentagelangspanish}{3.7\% }
\newcommand{\langfrench}{1.7\% }

\newcommand{\langchinese}{2.8\% }
\newcommand{\langportuguese}{0.3\% }
\newcommand{\langkorean}{0.1\% }

\newcommand{\percentagebiotrump}{3.25\% }
\newcommand{\percentagebiopatriot}{4.40\% }
\newcommand{\percentagebioconservative}{2.06\% }
\newcommand{\percentagebioreligious}{2.81\% }
\newcommand{\percentagebioqualifiers}{2.02\% }
\newcommand{\countverifiedusers}{1,609 }

\newcommand{\locbrazil}{5.41\% }
\newcommand{\loctexas}{0.79\% }
\newcommand{\locflorida}{0.52\% }
\newcommand{\loccalifornia}{0.26\% }
\newcommand{\locgermany}{0.24\% }
\newcommand{\loccanada}{0.16\% }
\newcommand{\locaustralia}{0.15\% }
\newcommand{\locfrance}{0.12\% }

\newcommand{\hashtagbiotrump}{0.38\% }
\newcommand{\hashtagbioqanon}{0.06\% }
\newcommand{\hashtagbiopatriot}{0.07\% }
\newcommand{\perbiourl}{4.60\% }
\newcommand{\perbiosocials}{21.18\%}
\newcommand{\perbiopornographic}{4.29\% }
\newcommand{\perbioblogs}{1.90\% }
\newcommand{\perverifiedusers}{0.18\%}

\newcommand{\followerstuckercarlson}{28,901 }
\newcommand{\followerstrumpjr}{24,716 }
\newcommand{\followersalexjones}{11,035 }

\newcommand{\perhttrump}{9.49\% }
\newcommand{\numhtbolsonaro}{5,185 }
\newcommand{\numhttransrightsarehumanrights}{1,719 }
\newcommand{\perhtcovid}{4.12\% }

\newcommand{\posturlyoutube}{17.57\%}

\newcommand{\urlbitchute}{1.39\%}
\newcommand{\urlodysee}{0.77\%}
\newcommand{\posturltwitter}{6.59\%}
\newcommand{\commenturltwitter}{4.65\%}
\newcommand{\urlrumble}{6.30\%}
\newcommand{\perhtbiden}{0.48\% }
\newcommand{\perverifiedfollowers}{29.34\%}

\begin{document}

\title{\bf\huge An Early Look at the Gettr Social Network}

\author{Pujan Paudel$^1$, Jeremy Blackburn$^2$, Emiliano De Cristofaro$^3$, Savvas Zannettou$^4$, and Gianluca Stringhini$^1$\\[0.5ex]
\normalsize{$^1$Boston University, $^2$Binghamton University, $^3$University College London, $^4$Max Planck Institute for Informatics}\\[0.25ex]
{\normalsize -- iDRAMA Lab, \url{https://idrama.science}} --
}

\date{}

\maketitle

\begin{abstract}
This paper presents the first data-driven analysis of Gettr, a new social network platform launched by former US President Donald Trump's team.
Among other things, we find that users on the platform heavily discuss politics, with a focus on the Trump campaign in the US and Bolsonaro's in Brazil.
Activity on the platform has steadily been decreasing since its launch, although a core of verified users and early adopters kept posting and become central to it.
Finally, although toxicity has been increasing over time, the average level of toxicity is still lower than the one recently observed on other fringe social networks like Gab and 4chan.
Overall, we provide a first quantitative look at this new community, observing a lack of organic engagement and activity.
\end{abstract}

\section{Introduction}

Over the past months, increased efforts by mainstream social networks to police their platforms and curb the spread of online harassment, misinformation, and conspiracy theories have led to unprecedented deplatforming and moderation actions.
As a consequence, a number of social media users have moved to alternative platforms.
This has led to a balkanization of the Web, where a myriad of smaller Web communities were created, where like-minded individuals can gather and keep discussing topics that may lead to suspension on mainstream platforms like Twitter.

In particular, we have seen the emergence of a number of alternative platforms that promise to promote free speech and allows their users to express themselves without fear of moderation.
Well-known examples include Parler~\cite{aliapoulios2021large,boyd2020parler,munn2021more} and Gab~\cite{fair2019shouting,zannettou2018gab}.
While these platforms do not have the audience of mainstream social networks like Twitter and Facebook, they constitute an important piece in the information ecosystem, and it is important for the research community to understand their influence in discussing news stories, popular events, as well as conspiracy theories and misinformation.

In this paper, we perform the first data-driven study of Gettr, the most recent of these alternative social platforms. 
During the month of July 2021 we collect 4.2M public posts and comments made by 419,273 users.
We then analyze these posts across several angles, from looking at the bios of Gettr users, to the activity of accounts over time, to their hashtag and URL posting activity.

In summary, we find that:

\begin{itemize}
\item Similar to other alternative platforms~\cite{aliapoulios2021large,zannettou2018gab}, discussion is dominated by topics related to the Trump campaign.
We also observe discussion on the Bolsonaro campaign in Brazil, on cryptocurrencies, and societal issues like the COVID-19 pandemic.
\item The influx of new users on Gettr seem to have quickly slowed down after its launch.
During July 2021 we find a steady decrease in user posting activity.
At the same time, verified users and early adopters keep posting on the platform, albeit at a reduced rate.
By the end of the observation period on July 31 2021, early adopters are responsible for 40\% of all daily posts and verified users of 5\% of posts.
 \item Although the level of (severe) toxicity in comments is increasing over time, it is still lower than the one observed by recent research on other alternative platforms, including Gab and 4chan~\cite{ali2021understanding,aliapoulios2021large}.
We also find that the level of identity attacks and threats on the platform is growing.
 \item Spam on Gettr is decreasing over time, which might be an indicator that the platform is deploying better anti-abuse mechanisms.
\end{itemize}

\section{What is Gettr?}
\label{sec:overview}

Gettr is a social network launched by former President Donald Trumps team, led by former Donald Trump aide and spokesman Jason Miller.
The platform went live on July 1, 2021 in its beta version, and was officially launched on July 4, 2021 for anyone to register.
The platform advertises itself as ``The  Marketplace of Ideas''  and lists itself as founded on the principles of, free speech, independent thought, and rejecting political censorship and ``cancel culture.''
The basic user interface and visuals of the  platform loosely follows the one of Twitter:
users can follow each other; and interact with others post by liking, reposting, and replying.
Gettr also uses the concepts of hashtags, highlighted hyperlinks in blue, allowing user to cluster posts related to a topic, and search posts using hashtags and trending hashtags.
The user profile section of Gettr is very close to the one of Twitter, where they allow users to upload their profile photos and banners respectively.
Similar to Twitter, user can also enter their bio, website, and location.
Posts, and replies on Gettr have the maximum length of 777 characters, and permit forms of media such as images, videos, GIFs and emojis.
Many popular accounts on Gettr are verified, however the formal process of verification is not explicitly mentioned in the platform.

According to the terms of service posted on the website, Gettr reserves the right, but is not committed to take down ``offensive, obscene, lewd, lascivious, filthy, pornographic, violent, harassing, threatening, abusive, illegal, or otherwise objectionable or inappropriate'' content.
On its launch day, it was reported that the platform had been compromised with some of the most prominent accounts of the platform being defaced~\cite{mashablearticle}.
There have also been security concerns regarding the platform, as hackers reportedly got hold of 90,000 Gettr user emails and locations~\cite{vicearticle}.
The platform was also reported to have been flooded with pornographic material during the initial days, which now seem to have been removed~\cite{insiderarticle}.

\section{Dataset}
In this section we describe our dataset.
First, we give a brief overview of Gettr's open, but undocumented, API.
Then we explain the methodology we used to collect 4.2M posts and comments from 419K users.
We also discuss limitations to with our collection methodology and ethical concerns with the collection.

\subsection{The Gettr API}

Although a full description of Gettr's API is beyond the scope of this paper, a quick overview is in order.
Before continuing, we note that Gettr's API is open, but undocumented.
That said, there have been many community efforts to reverse engineer and document the various endpoints Gettr exposes, e.g.,~\cite{gogettr}. In this paper, we make use of eight of these endpoints.

\descr{\texttt{/uinf}:} This endpoint is used to retrieve basic user information. E.g., their username, account creation date, user set location, language preferences, etc.

\descr{\texttt{/followers}:} This endpoint is queried to retrieve a list of users that follow a given user.

\descr{\texttt{/followings}:} This endpoint is queried to retrieve the list of users that this user follows.

\descr{\texttt{/posts}:} This endpoint is queried to retrieve the list of posts posted by a given user.

\descr{\texttt{/comments}:} This endpoint is queried to retrieve the list of comments under a given post.

\descr{\texttt{/liked}:} This endpoint is queried to retrieve the list of users who liked a given post.

\descr{\texttt{/shared}:} This endpoint is queried to retrieve the list of users who shared a given post.

\descr{\texttt{/post}:} This endpoint is queried to retrieve details about a given postid. The endpoint returns ``Content not found" when the post corresponding to the postid is not present.

\subsection{Collection Methodology}

We start our collection with user discovery, following a standard snowball sampling approach.
We seed our crawler with a list of eight users users in the ``Suggested for you'' page on Gettr.
Based on our preliminary analysis, this functionality returns a growing list of users, composed mostly of verified right-wing personalities and news media (newsmax,revolvernews,mikepompeo). 
To discover more users, we query the \texttt{/followers} and \texttt{/followings} endpoints for this set of seed users.
As new users are discovered, their followers and followings are also queried.
For each user we discover, we retrieve their posts and comments via the corresponding endpoints.
Like wise, for each comment or post we collect, we also query their \texttt{liked} and \texttt{shared} endpoints to get the list of users who liked, and shared the comments, and posts.

The primary limitation of our snowball sampling strategy is that we can only discover users that can be reached from our set of seed nodes.
We take another measure to explore any additional posts that might have been missed by the snowball sampling strategy.
We use the observation that identifier of posts (postids) on Gettr are base36 encodings of integers prefixed by a \texttt{p}.
We leverage the \texttt{/post} API on a range of integers (1 to 10 millions) to check if there are any mappings of valid postids that are absent in our dataset, and further crawl the entire timeline of the user, comments of all the user's posts, and the social network of the user who posted it. 

Another limitation with our strategy is that since we only crawl user timelines periodically we might miss posts and comments that are either deleted by the users themselves or by Gettr (e.g., spam content) between two successive recrawls.
Although we believe that our dataset is representative, we also suspect that getting a \emph{complete} dataset for Gettr will involve aggregating the results of on-going data collection efforts from other researchers.

The summary of the data collected is presented in  Table~\ref{table:statstable}.

\subsection{Ethical Considerations}
\jbnote{I need to jerk off a bit in here to make sure we don't have people crying that we are unethical.}
Collecting and analyzing social media data at scale has ethical implications.
In this work we only analyze data posted publicly and do not interact with users in any way.
As such, it is not considered human subjects research by the IRB at our institution.
Nonetheless, we adopt standard ethics guidelines to protect users~\cite{rivers2014ethical}.
More specifically, we only present aggregated data and we do not make efforts to further deanonymize the users in our dataset.

\begin{table}[t]
\center
\small
\begin{tabular}{@{}lrr@{}}
\toprule
\textbf{}         & \multicolumn{1}{c}{\textbf{Count}} & \multicolumn{1}{c}{\textbf{\#Users}} \\ \midrule
\textbf{Posts}    & 2,221,130                          & 334,901                              \\
\textbf{Comments} & 2,000,109                          & 236,734                              \\ \midrule
\textbf{Total}    & 4,221,239                          & 419,273                              \\ \bottomrule
\end{tabular}%
\caption{Overview of our dataset.}
\label{table:statstable}
\end{table}

\section{User Analysis}

We first look at the characteristics of Gettr users.
We are particularly interested in understanding the geographical distribution of the user base and their demographics, as well as the distribution of the number of followers and followings on the platform.
Finally, we are interested in understanding the daily posting and commenting activity on the platform.

We collect all user profile information for \totalusers Gettr accounts created between \datasetstart and \datasetend.
\inactiveusers accounts out of the total \totalusers made no posts or comments during our data collection period.
We also collect all posts, comments, and corresponding media posted by the accounts in our dataset until \datasetend.

\descr{User location and language.}
To understand the geographical makeup of Gettr users, we analyze the location and language from the bios of all Gettr users in our dataset.
Users can enter any arbitrary text as their location, which may not correspond to any physical location.
User languages are part of their profile settings where they can select languages to see posts, people, and trends in any language they choose.

Table~\ref{table:userloc} reports the most popular locations of the users in our dataset.
The most popular used location in user bios is Brazil, with \locbrazil of users (we combine ``Brazil'' and ``Brasil'' as a common location).
The next most popular locations are US states like Texas (\loctexas of users), Florida (\locflorida of users), and California (\loccalifornia of users).
We can also observe a few users from other countries such as Germany (\locgermany of users), Canada (\loccanada of users), Australia (\locaustralia of users), and France (\locfrance of users).

Table~\ref{table:userlang} reports the most popular language of the users in our dataset.
Despite Brazil being the most popular location, we find that the vast majority of Gettr users set English as their language (\percentagelangen of users), followed by Spanish (\percentagelangspanish of users).
The next most popular languages are Chinese (\langchinese of users) and French (\langfrench of users).
Portuguese and Korean only accounts for \langportuguese and \langkorean of the users, respectively.

\begin{table}[t]
\center
\small
\begin{tabular}{lrr}
\hline
\textbf{Location}       & \textbf{\#Users} & \textbf{\%Users} \\ \midrule
Brazil         & 46,325  & 5.41\%   \\
Texas          & 6,838   & 0.79\%   \\ 
USA            & 6,451   & 0.75\%   \\ 
Florida        & 4,521   & 0.52\%   \\ 
California     & 2,276   & 0.26\%   \\ 
Germany        & 2,119   & 0.24\%   \\ 
São Paulo      & 2,113   & 0.24\%   \\ 
Rio de Janeiro & 1,837   & 0.21\%   \\ 
Arizona        & 1,617   & 0.18\%   \\ 
Ohio           & 1,490   & 0.17\%   \\ 
Michigan       & 1,437   & 0.16\%   \\ 
Canada         & 1,403   & 0.16\%   \\ 
Australia      & 1,363   & 0.15\%   \\ 
New York       & 1,247   & 0.14\%   \\ 
Georgia        & 1,216   & 0.14\%   \\ 
Tennessee      & 1,114   & 0.13\%   \\ 
North Carolina & 1,103   & 0.12\%   \\ 
France         & 1,034   & 0.12\%   \\ 
Alabama        & 934     & 0.10\%   \\ \bottomrule
\end{tabular}
 \caption{Top 20 user locations based on user bios.}\label{table:userloc}
\end{table}

\begin{table}[t]
  \center
\begin{tabular}{lrr}
\hline
\textbf{Language} & \textbf{\#Users} & \textbf{\%Users} \\ \midrule
English       & 767,589 & 90.19\% \\ 
Spanish       & 31,811  & 3.71\%  \\ 
Chinese   & 24,117  & 2.81\%  \\ 
French       & 14,984  & 1.75\%  \\ 
Japanese       & 7,552   & 0.88\%  \\ 
Portuguese   & 2,887   & 0.33\%  \\ 
Korean       & 1,094   & 0.12\%  \\ \bottomrule
\end{tabular}
\caption{Popularity of languages in our Gettr users.}
\label{table:userlang}
\end{table}

\descr{User bios.}
Next, we analyze the language included by Gettr users in their bios.
To this end, we extract the most popular words and bigrams used in bios.
Table~\ref{tabel:userbios} reports the top keywords and bigrams observed in user bios in our dataset.
As it can be seen, a large fraction of users include keywords related to former President Trump and his campaign, indicated by users supporting him (\percentagebiotrump of all users included the term ``trump,'' ``trump supporter,'' ``trump won,'' ``president trump,'' ``maga'').
We also find that many users self identify as patriots (\percentagebiopatriot of the users use ``patriot,'' ``american,'' ``proud,'' ``country''), conservatives (\percentagebioconservative of the users use ``conservative,'' ``conservador,'' ``christian conservative''), and religious (\percentagebioreligious of the users use ``god,'' ``christian,'' ``crist'').
We also observe \percentagebioqualifiers of users using personal qualifiers such as retired, father, wife, mother in their bios.
These results suggest that the user base of Gettr is very similar to the one of other alternative social platforms like Gab and Parler~\cite{zannettou2018gab,aliapoulios2021large}.

Table~\ref{table:userbiosextra} lists the top hashtags included by Gettr users in their bios.
As it can be seen, the use of hashtags in bios is dominated by hashtags supporting former President Trumps campaign ( \hashtagbiotrump  of users using \#MAGA , \#TrumpWon, \#Trump2024, \#KAG).
We can also see instances of the QAnon conspiracy theory~\cite{aliapoulios2021gospel,papasavva2020qoincidence,phadke2021characterizing} (\hashtagbioqanon  of users using \#WWG1WGA), and hashtags indicating  patriotic sentiments (\hashtagbiopatriot  of users using \#AmericaFirst, \#SaveAmerica, \#Patriot).

Gettr users can also add website in their user bios, either by including a URL in the bio field, or through a separate field called ``website.''
Table~\ref{tabel:userbios} lists the most popular domains included in user bios in our dataset.
\perbiourl of all the users include a website or link in their bio, with \perbiosocials of users include links to other social media, including YouTube, Instagram, Twitter, and Facebook.
We can also see traces of pornographic websites on user bios \perbiopornographic.
We also observe \perbioblogs users including links to personal blogs (blogspot, wordpress, wix). %

\begin{table}[t]
\small
\center
\begin{tabular}{lrlr}
\hline
\textbf{Word}         & \textbf{\#Users} & \textbf{Bigrams}               & \textbf{\#Users} \\ \midrule
Trump        & \multicolumn{1}{r|}{15,754}    & trump supporter        & 2,296     \\ 
patriot      & \multicolumn{1}{r|}{14,009}    & trump won              & 2,251     \\ 
love         & \multicolumn{1}{r|}{11,829}    & husband father         & 1,778     \\ 
god          & \multicolumn{1}{r|}{11,428}    & proud american         & 1,771     \\ 
conservative & \multicolumn{1}{r|}{11,282}    & deus acima             & 1,679     \\ 
american     & \multicolumn{1}{r|}{10,700}    & american patriot       & 1,667     \\ 
christian    & \multicolumn{1}{r|}{7,617}     & america first          & 1,550     \\ 
america      & \multicolumn{1}{r|}{7,576}     & god family             & 1,547     \\ 
proud        & \multicolumn{1}{r|}{6,706}     & my country             & 1,446     \\ 
retired      & \multicolumn{1}{r|}{6,421}     & president trump        & 1,414     \\ 
freedom      & \multicolumn{1}{r|}{6,419}     & she her                & 1,388     \\ 
country      & \multicolumn{1}{r|}{6,267}     & brasil acima           & 1,358     \\ 
maga         & \multicolumn{1}{r|}{6,135}     & christian conservative & 1,307     \\ 
family       & \multicolumn{1}{r|}{5,415}     & de direita             & 1,304     \\ 
conservador  & \multicolumn{1}{r|}{5,103}     & trump is               & 1,298     \\ 
crist        & \multicolumn{1}{r|}{5,057}     & god bless              & 1,280     \\ 
father       & \multicolumn{1}{r|}{4,997}     & the truth              & 1,271     \\ 
life         & \multicolumn{1}{r|}{4,970}     & wife mother            & 1,251     \\ 
wife         & \multicolumn{1}{r|}{4,646}     & free speech            & 1,163     \\ \bottomrule
\end{tabular}
\caption{Top 20 words and bigrams on user bios.}
\label{tabel:userbios}
\end{table}

\begin{table}[t]
\centering
\begin{tabular}{lrlr}
\hline
\textbf{Hashtag}       & \textbf{\#Users} & \textbf{Domain}             & \textbf{\#Users} \\ \midrule
MAGA          & \multicolumn{1}{r|}{1,963}    & youtube.com        & 3,754    \\ 
TrumpWon      & \multicolumn{1}{r|}{683}     & instagram.com      & 1,963    \\ 
WWG1WGA       & \multicolumn{1}{r|}{565}     & twitter.com        & 1,733    \\ 
2A            & \multicolumn{1}{r|}{532}     & sexualpartner3.com & 1,154    \\ 
Trump2024     & \multicolumn{1}{r|}{374}     & facebook.com       & 900     \\ 
AmericaFirst  & \multicolumn{1}{r|}{296}     & t.me               & 842     \\ 
KAG           & \multicolumn{1}{r|}{243}     & linktr.ee          & 826     \\ 
1A            & \multicolumn{1}{r|}{216}     & blogspot.com       & 355     \\ 
Bolsonaro2022 & \multicolumn{1}{r|}{207}     & turnmeon.com      & 313     \\ 
SaveAmerica   & \multicolumn{1}{r|}{205}     & gettr.com          & 301     \\ 
wap           & \multicolumn{1}{r|}{170}     & discord.com        & 291     \\ 
Patriot       & \multicolumn{1}{r|}{165}     & supervideos.online & 275     \\ 
Prolife       & \multicolumn{1}{r|}{109}     & pornhub.com        & 227     \\ 
NRA           & \multicolumn{1}{r|}{98}      & tiktok.com         & 218     \\ 
BackTheBlue   & \multicolumn{1}{r|}{91}      & wixsite.com        & 203     \\ 
Conservative  & \multicolumn{1}{r|}{80}      & wordpress .com     & 192     \\ 
Christian     & \multicolumn{1}{r|}{79}      & gab.com            & 192     \\ 
GodWins       & \multicolumn{1}{r|}{78}      & linkedin.com       & 168     \\ \bottomrule
\end{tabular}
\caption{Top 20 hashtags and domains on user bios.}
\label{table:userbiosextra}
\end{table}

\descr{Post and comment activity.}
As discussed in Section~\ref{sec:overview}, users on Gettr can get a verification badge, similar to what happens on other social networks.
In this section we compare the post and commenting activity of general Gettr users compared to verified ones.
To this end, we extract \countverifiedusers users in our dataset that are verified by Gettr (\perverifiedusers).

\begin{figure}[t]
\includegraphics[width=0.9\columnwidth]{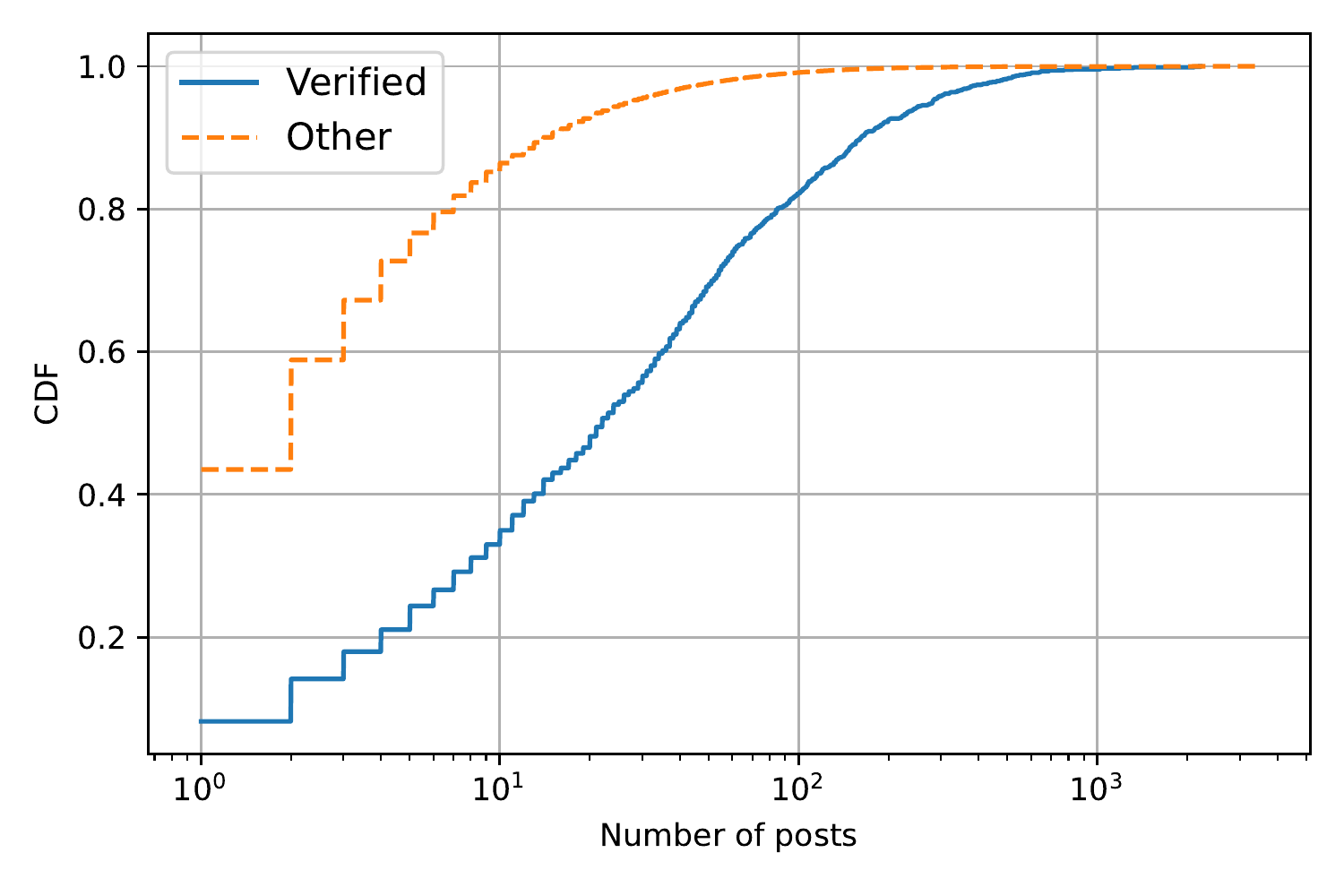}
\caption{CDF of the number of posts made by users. Note the log scale for the x-axis.}
\label{fig:ccdf_posts}
\end{figure}

We plot the cumulative distribution function (CDF) of the number  posts per user split by user type (verified, normal) in Figure~\ref{fig:ccdf_posts}.
In general, verified users are more active posting on the platform than the non-verified users.
More than 80\% of the unverified Gettr users have less than 10 posts, which shows that a large chunk  of accounts exist on the platform with little to no activity.
Verified users engage more in posting behavior than unverified users, but not at the frequency that we would expect, as almost 80\% of the verified users post at most 100 times during our observation period.
Still, there are more verified users who have between 100-1000 posts, compared to the very skewed distribution of unverified users posting in the same range.
We run a two sample Kolmogorov-Smirnoff (KS) test~\cite{kstest} on the two CDFs, and find that the differences between the difference in the distribution of posts and comments between the two user types is statistically significant at the $p < 0.001$ level.

\begin{figure}[t]
\includegraphics[width=0.9\columnwidth]{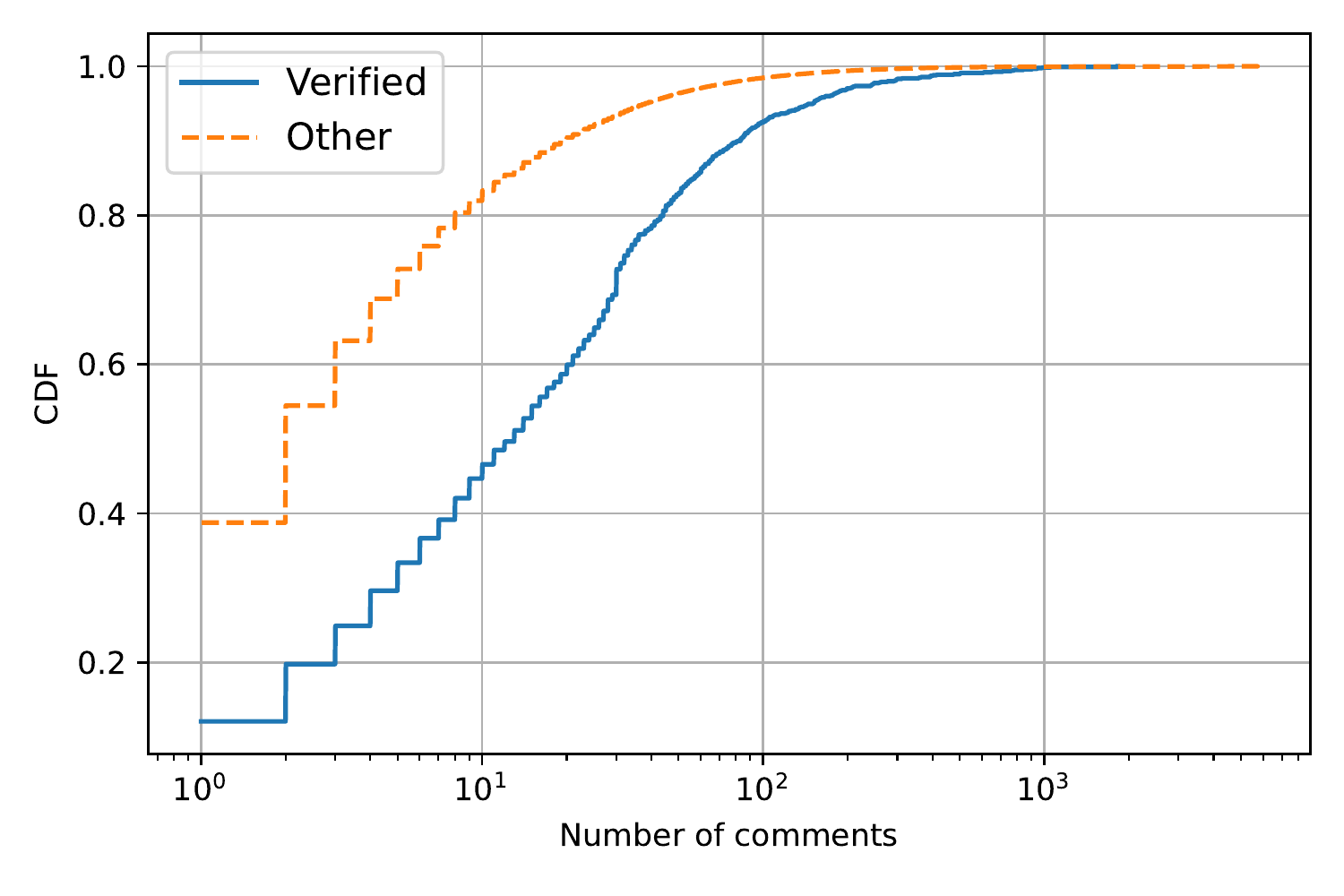}
\caption{CDF of the number of comments made by users. Note the log scale for the x-axis.}
\label{fig:ccdf_comments}
\end{figure}

We observe similar patterns in the case of commenting behavior in Figure~\ref{fig:ccdf_comments}.
However, while verified users tend to post more than comment, the opposite is true for regular users.
Close to 90\% of the verified users have less than or equal to 100 comments. %
In the case of unverified users, close to 40\% of the users comment between 3-10 times.
Again, a two sample Kolmogorov-Smirnoff (KS) test finds that the distribution of comments between verified and regular users shows statistically significant differences at the $p < 0.001$ level.

\begin{figure}[t]
\includegraphics[width=\columnwidth]{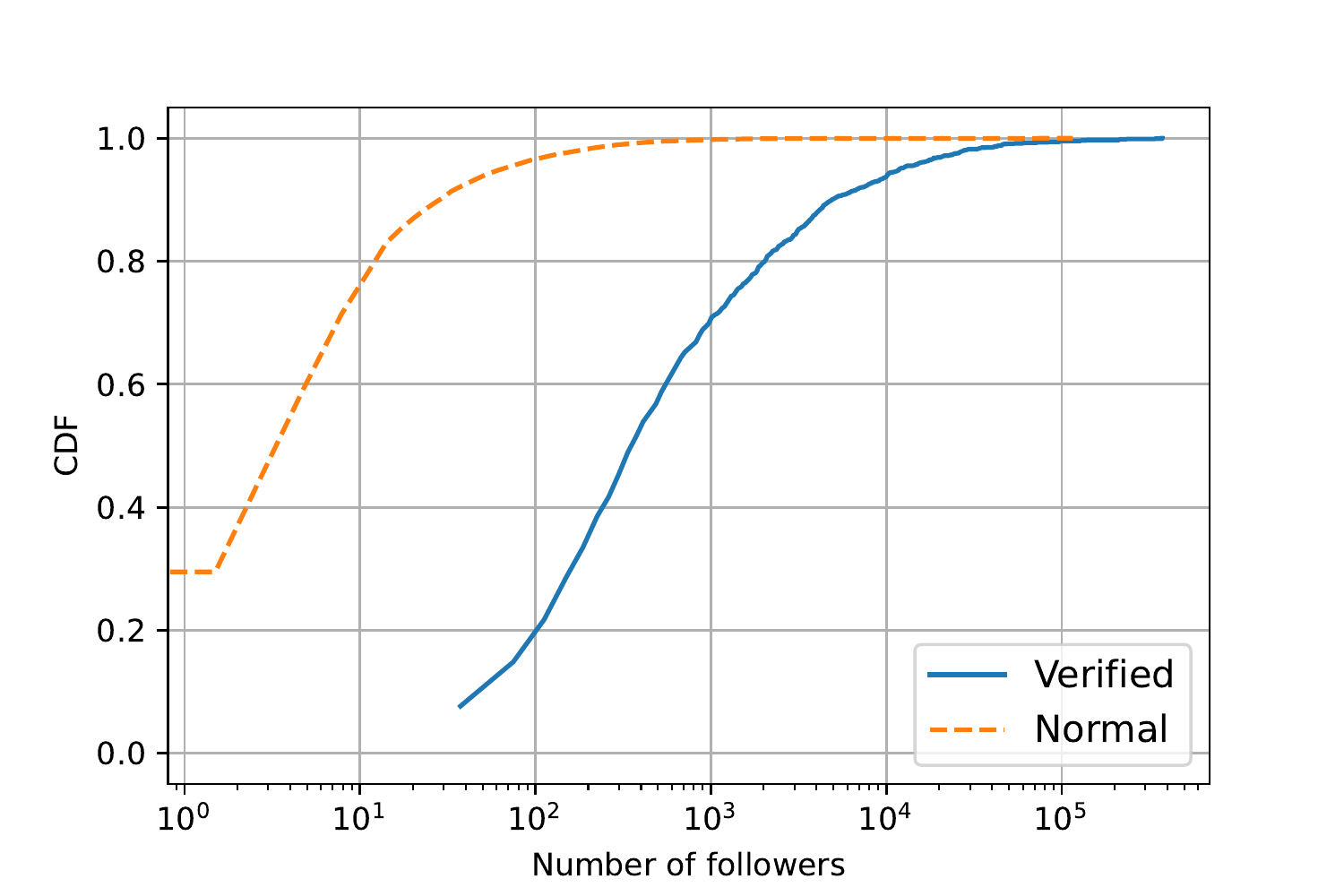}
\caption{CDF of the number of followers of users. Note the log scale for the x-axis.}
\label{fig:ccdf_followers}
\end{figure}

\begin{figure}[t]
\includegraphics[width=\columnwidth]{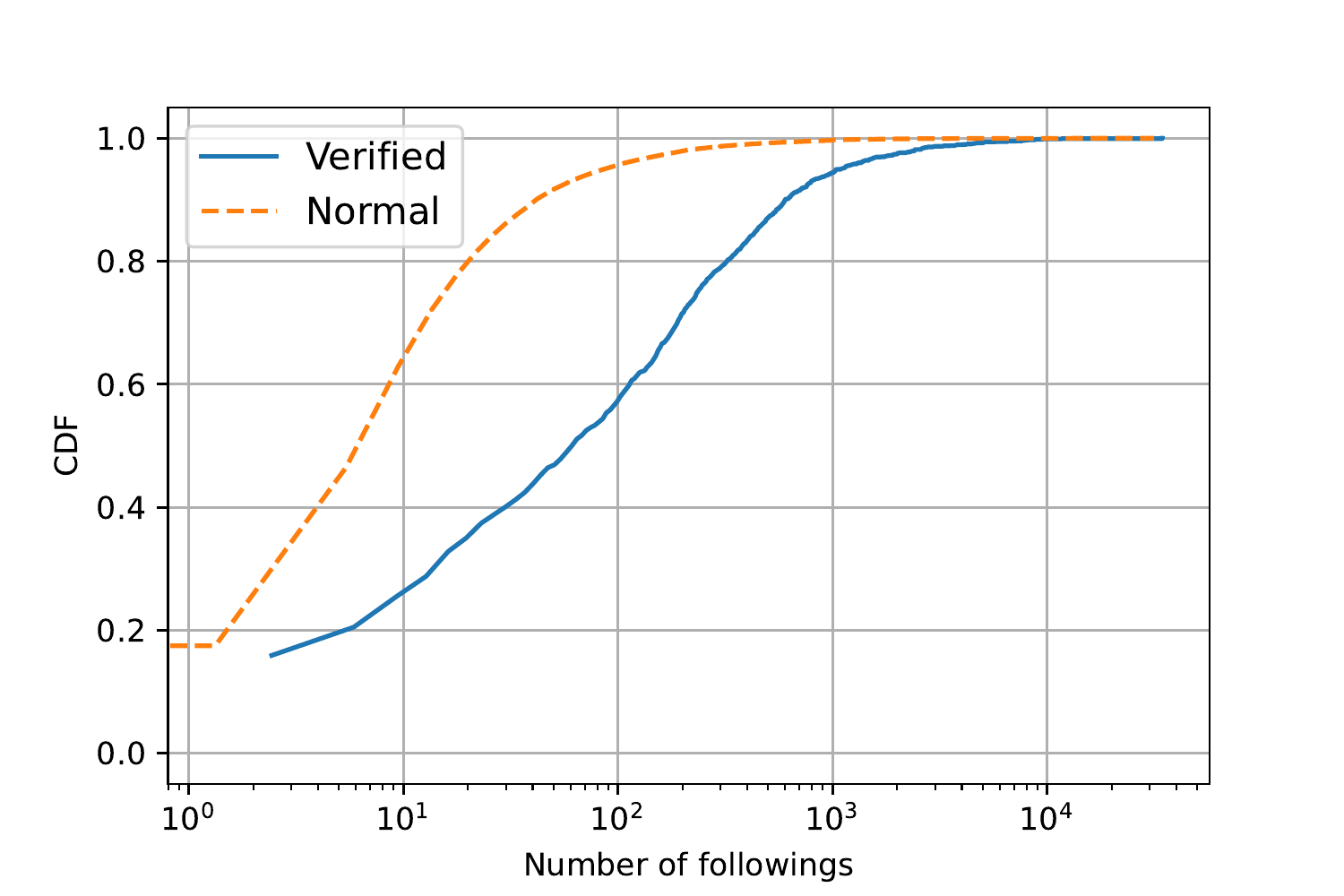}
\caption{CDF of the number of followings of users. Note the log scale for the x-axis.}
\label{fig:ccdf_followings}
\end{figure}

\descr{User followers/followings.}
Next, we look at the number of accounts that users followed or are followed by on Gettr.
To this end, we plot the cumulative distribution function (CDF) of the  followers count per users, split by verified or unverified status in Figure~\ref{fig:ccdf_followers}.
As it can be seen, a large fraction (close to 80\%) of the unverified users have less than 10 followers, and more than 95\% of the users have less than 100 followers. %
On the other hand, verified users garner a good number of followers, with \perverifiedfollowers of the verified users having more than a thousand followers.
Upon further investigation, the only two unverified user accounts with more than 100,000 followers had usernames \texttt{presdonaldtrump}, and \texttt{trumpteam}, most likely impersonating Donald Trump and his team.
The accounts are now suspended at the time of writing.
The other unverified users who have a high number of followers (\texttt{carlazambelli} with 87,107, and  \texttt{canalhipocritas} with 37,750) are related to Brazilian politics.
We also found other accounts trying to impersonate popular right-wing personalities like Tucker Carlson (\followerstuckercarlson followers), Donald Trump Jr. (\followerstrumpjr followers), and Alex Jones (\followersalexjones followers), some of which have now been suspended.

The CDF of the following distributions of users is plotted in Figure~\ref{fig:ccdf_followings}.
As seen in most of the social media platforms, the majority of verified users follow less people than they are followed by~\cite{cha2010measuring}.
We also find that there are five verified users who follow more than eight thousand users, which is a relatively high number of followings for a verified user.
These were found to be anti-CCP news outlets posting their content in Chinese.
Following the similar Kolmogorov-Smirnoff (KS) test we did for posts and comments, we find that the differences between the distributions of followers and followings between the two user types are statistically significant at the $p < 0.001$  level.

\begin{figure}[t]
\includegraphics[width=\columnwidth]{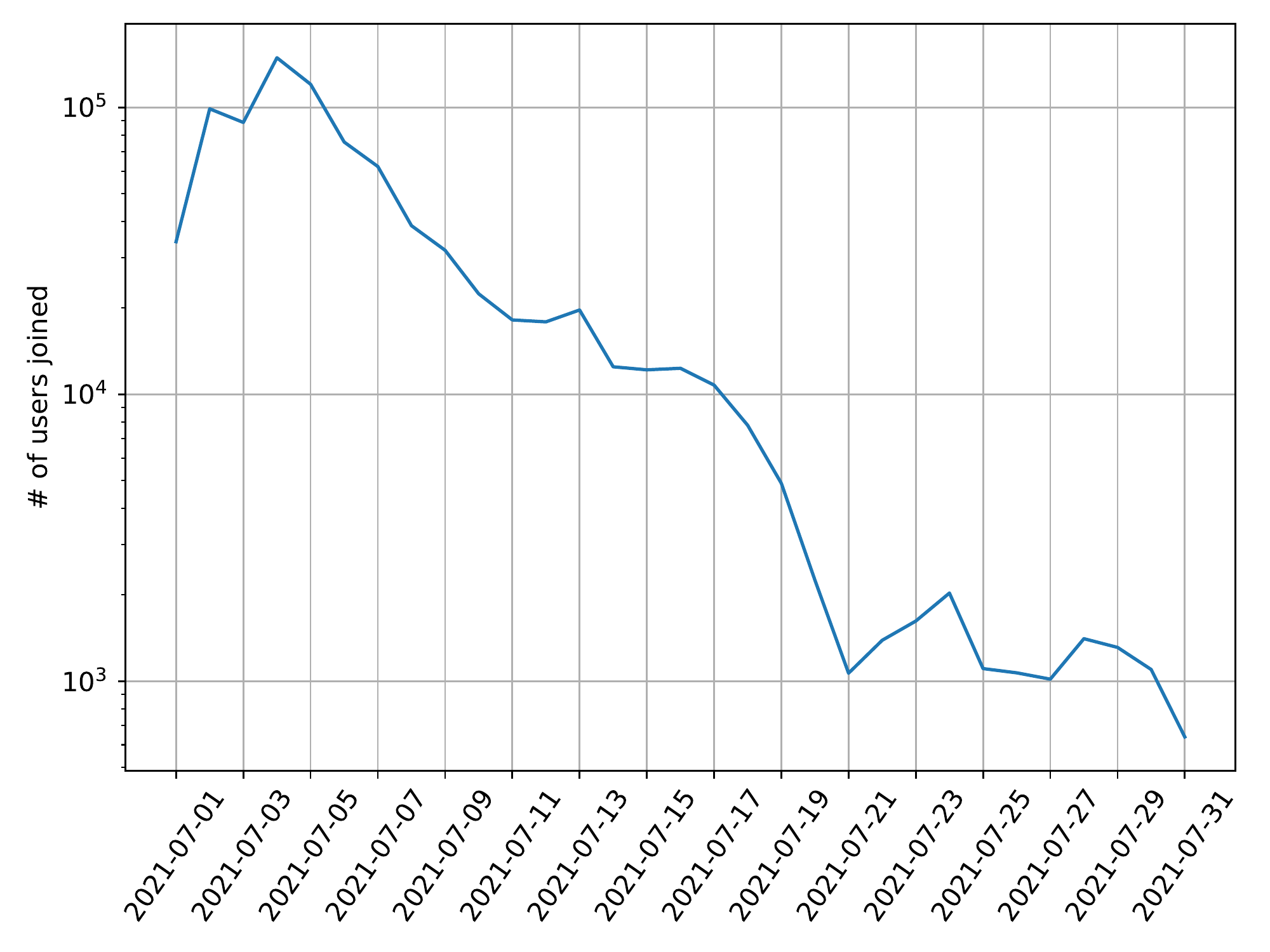}
\caption{Evolution of number of unique users joining the platform. Note the log scale on x-axis.}
\label{fig:creationdate}
\end{figure}

\descr{User creation date.}
The Gettr platform appeared online first at July 1, and was officially launched on July 4.
In this section we look at when the accounts in our dataset were created, to get a sense of the user influx on Gettr.
Figure~\ref{fig:creationdate} shows the number of accounts created daily during our observation period.
As it can be seen, a large number of accounts were created before July 4 (34,897 in July 1, 101,281 in July 2, and 90,403 in July 3) despite the platform only being in Beta.
The number of users joining the platform increases and peaks on July 4 (151,598), the day of official release and July 5 (122,587), the day after.
We can observe that the number of users joining the platform has been on the decline since the official launch, however the number of users joining the platform has been in the order of thousands per day. %

\begin{figure}[t]
\includegraphics[width=\columnwidth]{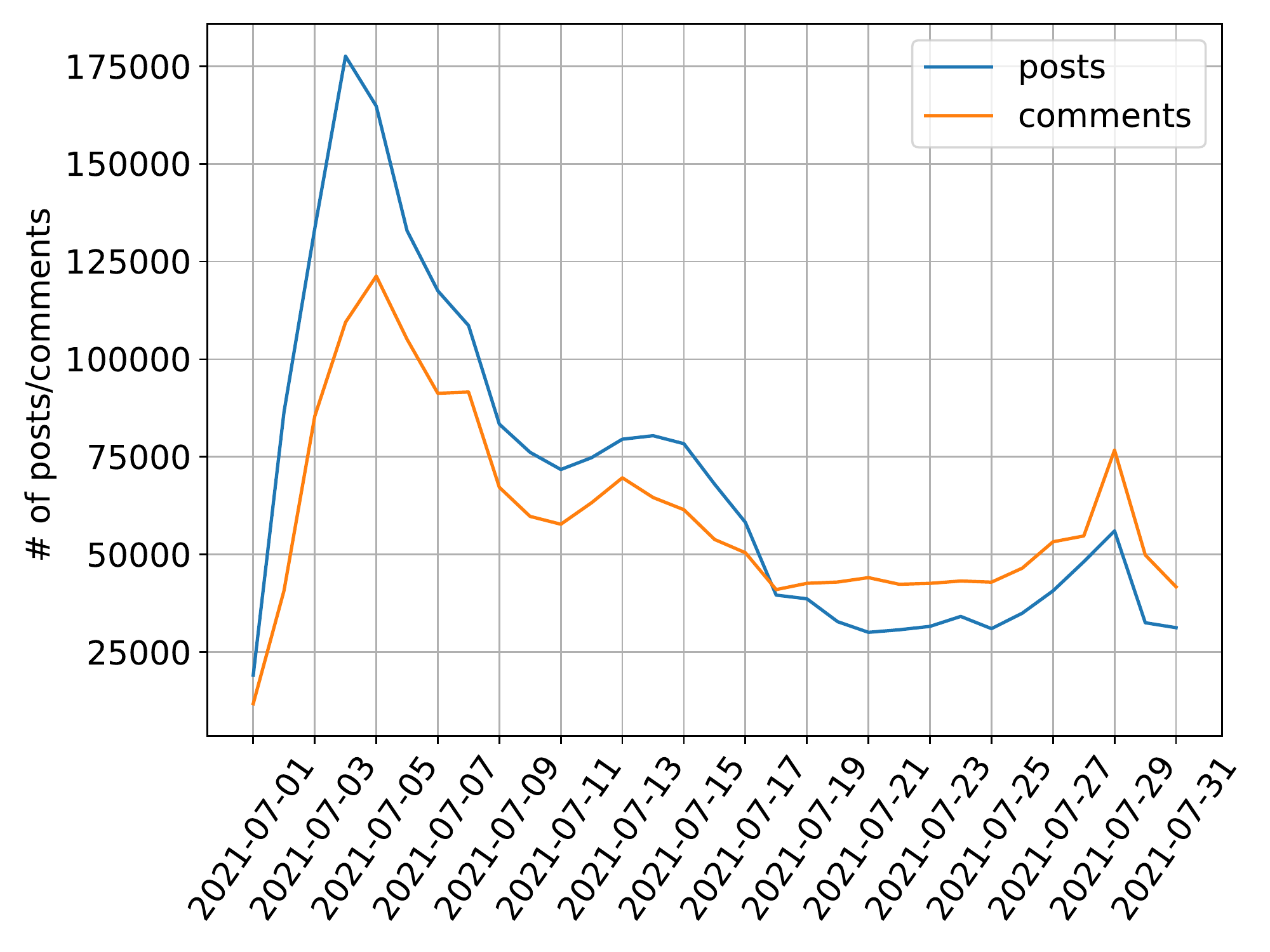}
\caption{Evolution of user activity.}
\label{fig:activityjuly}
\end{figure}

\descr{Daily user activity.}
Next we are interested in looking at the daily activity of users on Gettr.
Figure~\ref{fig:activityjuly} shows the progression of the daily number of posts and comments observed in our dataset for the month of July 2021.
As it can be seen, as the official release date approached, the number of daily posts and comments started increasing in the platform.
The daily number of posts peaks on the official date of release (July 4) with 195,229 posts on the day, while the number of comments peak a day later (July 5) with 142,104 comments.
After the peak date, we can see a very steep decline on the daily number of new posts and comments (180,718, and 144,391 respectively following the peak date for the posts while 119,727, and 104,961 for the comments).
Eventually, there seems to be saturation towards creation of posts as there are more daily comments than posts starting from July 18, and the trend seems to be consistent after that day.

Our dataset also has 7,275 posts and 6,565 comments that appeared before the platform went live on July 1, which are most likely content automatically imported from Twitter.
The posts and comments imported from Twitter are mostly concentrated near the days close to July 1, as Twitter only allows retrieval of the past 3,200 tweets of a user profile.
The importing of tweets functionality was blocked by Twitter on July 10~\cite{twitterimport}.

The general trend of user activity indicates that as time progresses the number of posts on Gettr strictly declined, while the number of comments stabilized.
To get a more nuanced picture of user activity on Gettr, we focus on two sets of users: those who have been verified by the platforms and those that were early adopters of the platform, joining on the official launch day (July 4, 2021) or earlier.

\begin{figure}[t]
\includegraphics[width=\columnwidth]{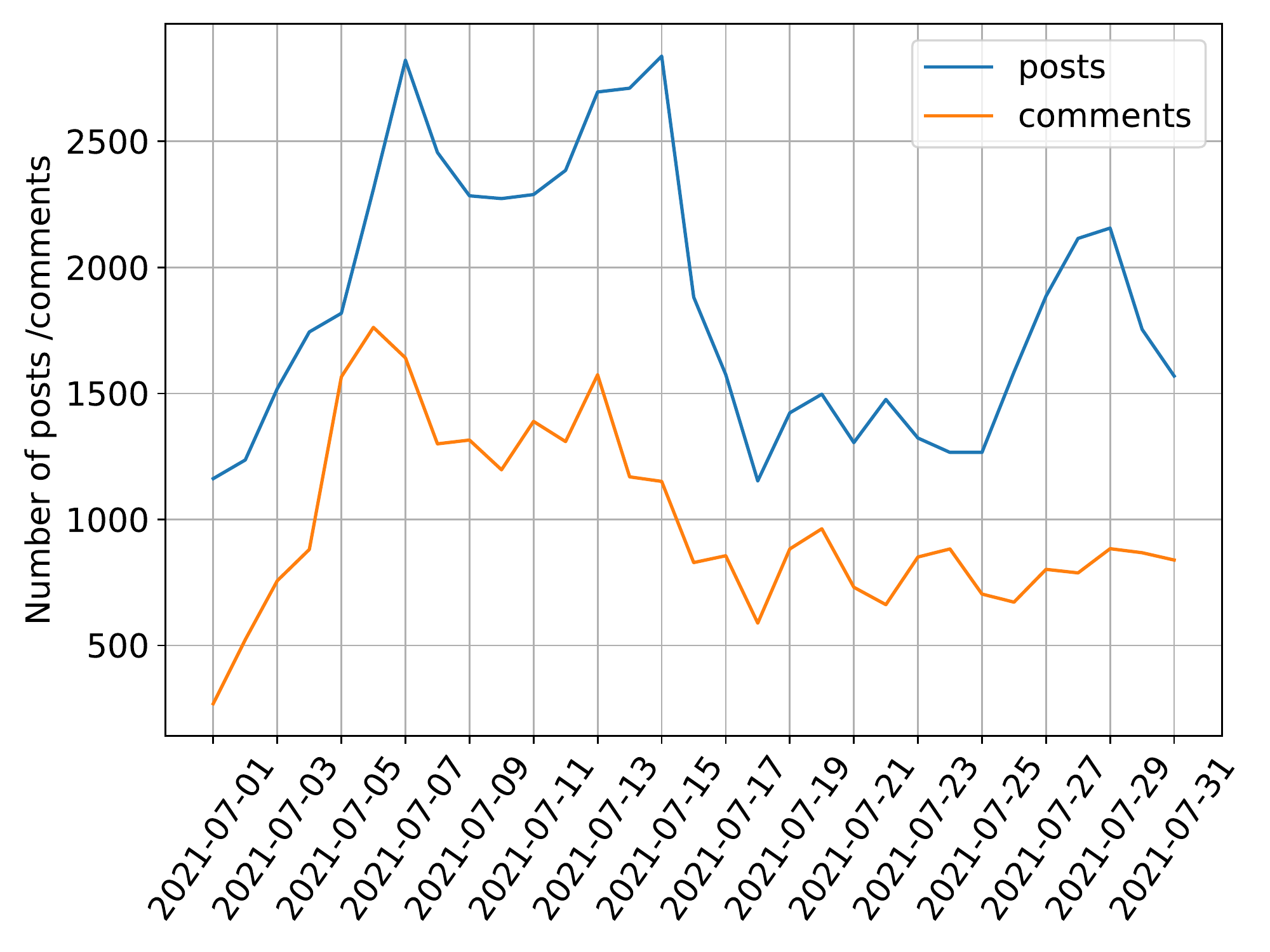}
\caption{Raw volume of verified users' activity.}
  \label{fig:verifiedactivity_raw}
\end{figure}

\begin{figure}[t]
\includegraphics[width=\columnwidth]{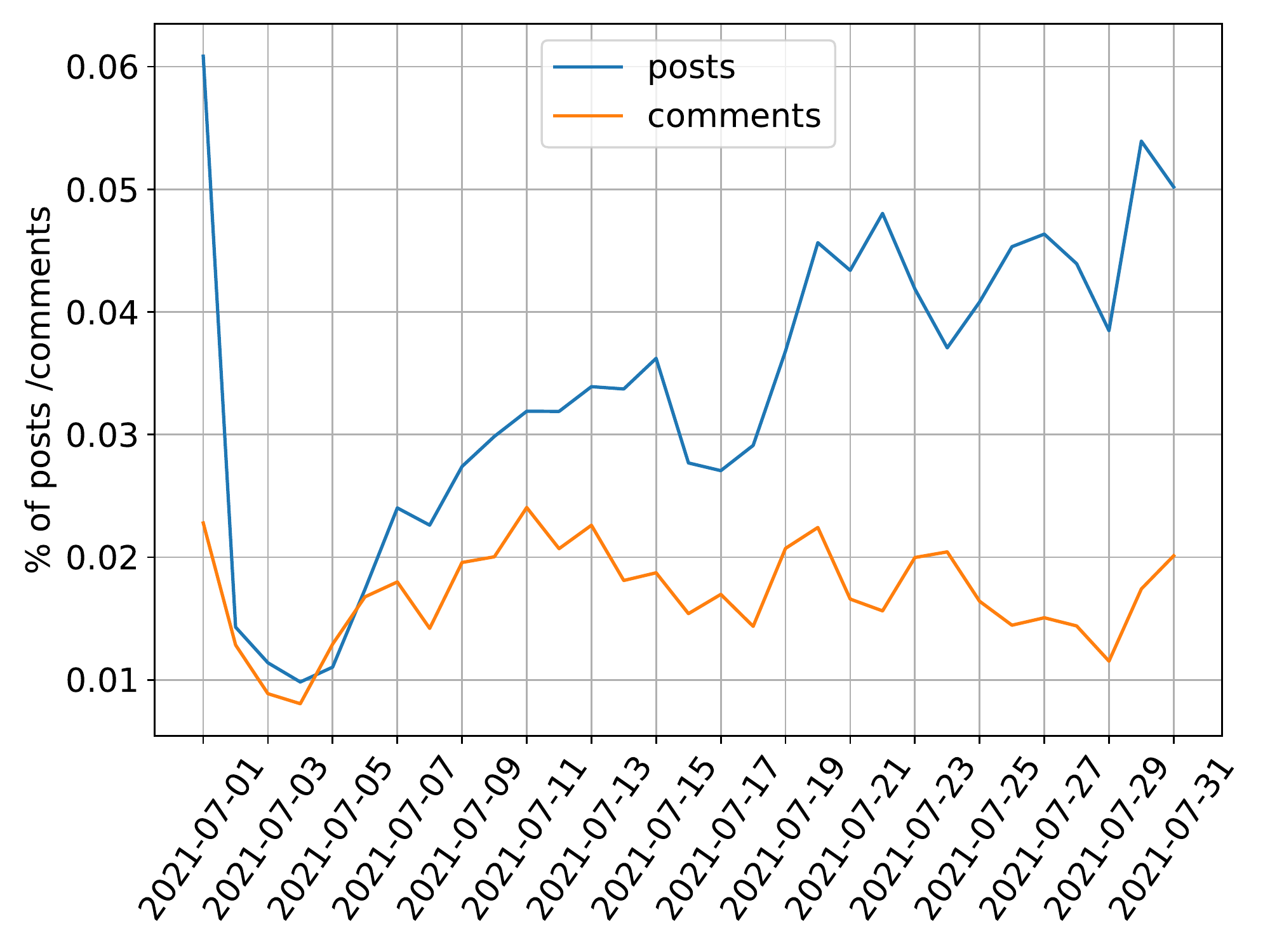}
\caption{Percentage contribution of verified users' activity.}
\label{fig:verifiedactivity_perc}
\end{figure}

Recall that there are \countverifiedusers verified users in our dataset.
Figures~\ref{fig:verifiedactivity_raw} and~\ref{fig:verifiedactivity_perc} show the time progression of the number of comments and posts and their relative percentage by the verified users on each day.
On July 1, verified users were responsible for 6\% (close to 1,200) of all posts on Gettr.
As we can see from Figure~\ref{fig:verifiedactivity_raw}, the decline in posting activity by verified users is not as pronounced as the one by the general population, and it stabilizes around 1.5K-2K daily posts after July 18.
Because of this, the relative percentage of posts made by verified users shows an increasing trend, stabilizing at around 5\% daily posts (see Figure~\ref{fig:verifiedactivity_perc}).
This indicates that verified users keep being active and making posts on Gettr throughout the observation period.
When looking at comments, on the other hand, verified users are responsible for 1-2\% of daily comments throughout most of the observation period, with a mostly stable trend.
The difference between the trends of the verified users on posts and comments contribution shows that verified users focus more on creating original content rather than commenting on existing one.

\begin{figure}[t]
\includegraphics[width=\columnwidth]{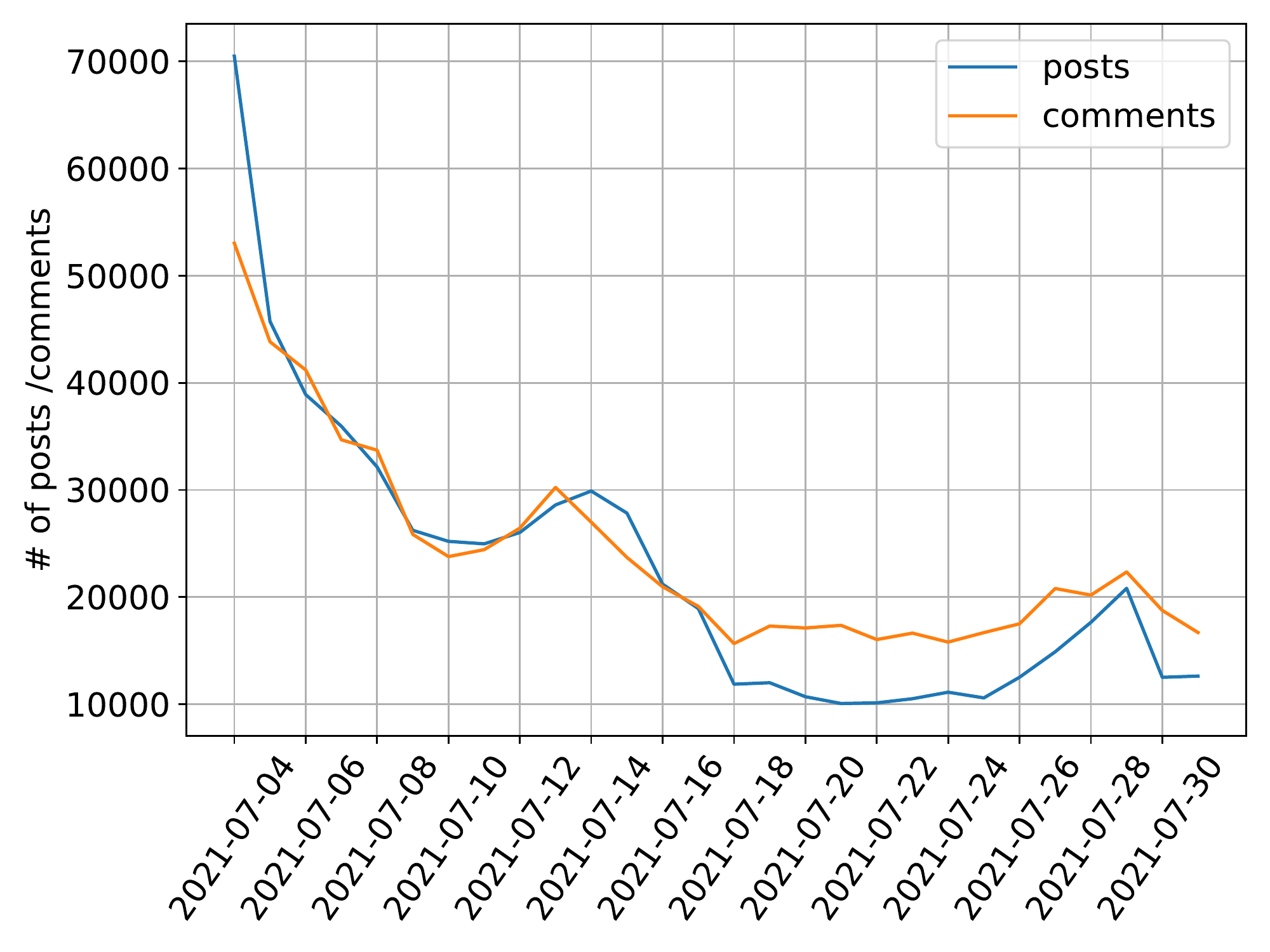}
\caption{Raw volume of early adopters' activity.}
\label{fig:earlyadopteractivity_raw}
\end{figure}

\begin{figure}[t]
\includegraphics[width=\columnwidth]{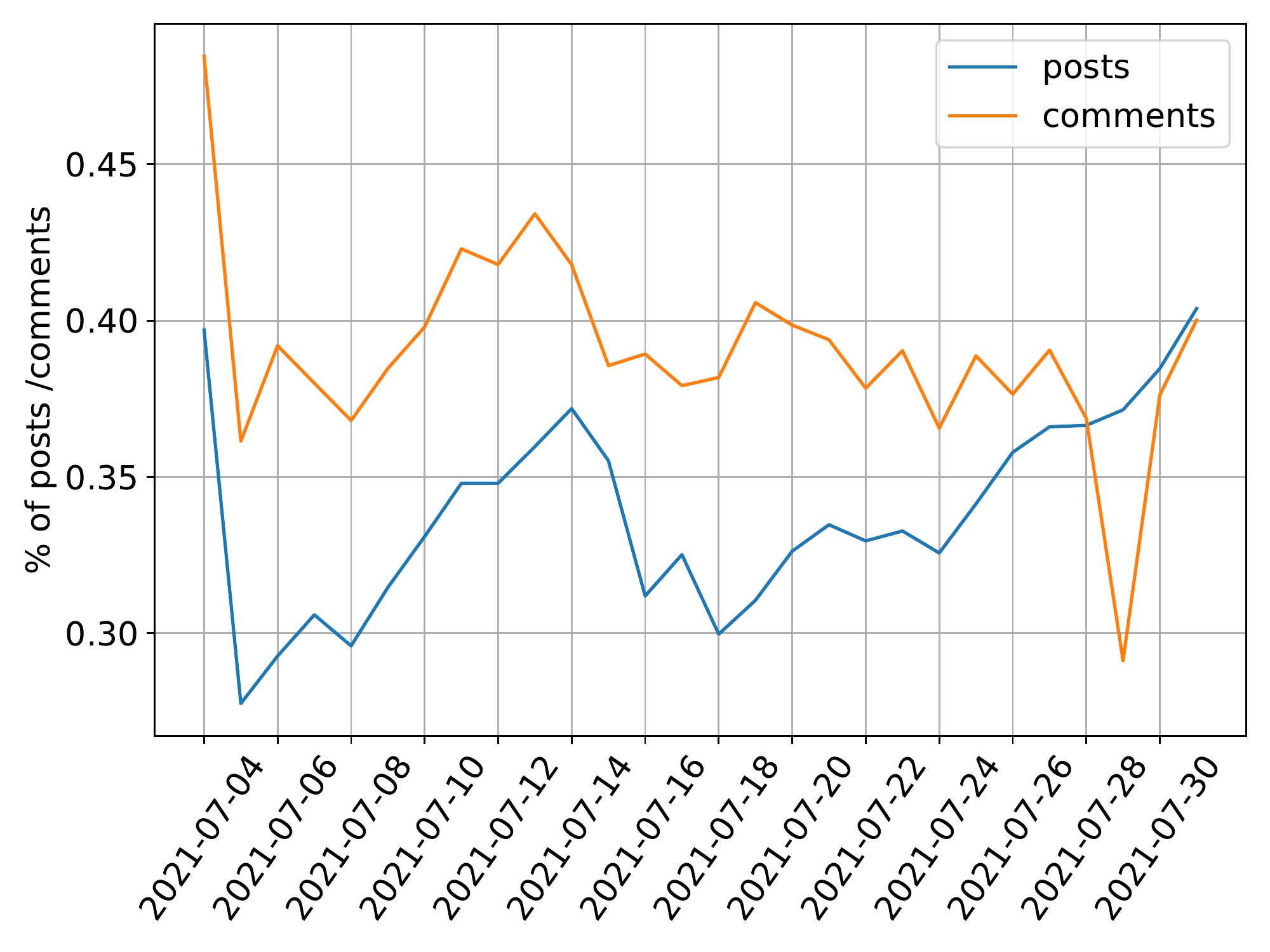}
\caption{ Percentage contribution of early adopters' activity.}
\label{fig:earlyadopteractivity_perc}
\end{figure}

When looking at early adopters, we find that 226,581 users signed up on Gettr before the official launch day of July 4.
Figures~\ref{fig:earlyadopteractivity_raw} and~\ref{fig:earlyadopteractivity_perc} show the raw volumes and percentage of daily posts and comments that early adopters were responsible for on Gettr.
As it can be seen in Figure~\ref{fig:earlyadopteractivity_raw}, early adopters experiences a similar decrease in posting activity than the general user population, declining from 70,000 posts a day to 25,000 posts a day.
When looking at percentages from Figure~\ref{fig:verifiedactivity_perc}, however, we can observe that early adopters contribute a large share of discussion every day (between 30-45\%) throughout the month of July.
We also observe that the fraction of posts contributed by early adopters steadily increases after July 18, suggesting that the platforms user base might be crystallizing on this group of early adopters.
Towards the later part of the month, the early adopters start commenting more (18,000 daily posts)  than they are posting (12,000 daily posts), signaling a shift in user behavior.
Differently from verified users, early adopters account for a higher fraction of comments than they do of posts.

\section{Content Analysis}
Gettr bills itself as a place for ``unbiased'' discussion, in the same vein as Parler and Gab before it.
A major differentiating factor with Gettr is that it has direct ties to former, high ranking members of the Trump administration (along with alleged ties to Chinese billionaire, Guo Wengui~\cite{guofundinggettr}).
Although a deeper exploration of the underlying organization and business model of Gettr is out of scope of this paper, we note that these ties to known political actors is a strong indicator that Gettr is \emph{not} unbiased.

With the above said, in this section we analyze the content posted to Gettr along several axes.
First, we take a brief look at the most popular sites linked to in Gettr posts and comments.
Because Gettr is in large part a functional clone of Twitter, we also explore the use of hashtags.
Finally, we measure how toxicity and spam has evolved on Gettr over time.

\descr{Hashtags.}
As a first step to understanding the conversation on Gettr, we look at the most popular hashtags.
Previous explorations of Gab and Parler via hashtags have made it clear that their user base had a clear right-wing bias, with Trump related hashtags dominating the discussion~\cite{aliapoulios2021large,zannettou2018gab}, and this mostly holds for Gettr as well.

Table~\ref{table:postshashtags} lists the top 20 hashtags that appeared in posts within our dataset.
  \texttt{\#covid} was the most popular hashtag in posts, accounting for \perhtcovid of all hashtags.
  \perhttrump of the hashtags used in posts are Trump related (\texttt{\#trumprally}, \texttt{\#trumpwon}, \texttt{\#maga}, \texttt{\#trump2024}), specifically pushing ``The Big Lie~\cite{bigliecnn},'' with \texttt{\#biden} accounting for another \perhtbiden of the total hashtags.

Interestingly, \numhtbolsonaro posts used \texttt{\#bolsonaro2022}, which fits with the over 5.4\% of accounts reporting their location as being in Brazil.
There are several potential explanations for this.
One explanation is that the contingent of Brazilian accounts are legitimate and aligned with Bolsonaro's politics.
Another, more cynical explanation is that these are not legitimate accounts and are involved in anything ranging from a simple spam campaign to a more sophisticated operation a-la state sponsored social media trolls~\cite{badawy2018analyzing,starbird2018ecosystem,zannettou2019let}.

We also observe a few unexpected hashtags in posts.
  For example \texttt{\#didi} is referring to the Chinese ride-hailing app that acquired Uber China in 2016, and users on the platform seem to be criticizing the removal of the app from App store~\cite{didiremoval}.

  We also observe hashtags like \texttt{\#heatwave} and \texttt{\#mudslide}, which appear to have been initially been related to climate/natural disaster discussion and were later hijacked and used in completely unrelated contexts, such as promotion of Korean-pop artists, and used alongside images of Arabic scripts with other unrelated hashtags (\texttt{\#ElonMusk} and \texttt{\#bitcoin}).

\jbnote{Write something about bitcoin}

\jbnote{END NEW STUFF}

Table~\ref{table:commentshashtags} lists the top 20 hashtags that appeared in comments within our dataset.
Although the overall theme of the most popular hashtags used in comments is the same as those used in posts, there are some notable differences.
  For example, the QAnon related hashtag \texttt{\#wwg1wga} appears in 844 comments.
  We also see some hashtags clearly related to The Big Lie, but here some additional nuance comes into the picture with \texttt{\#trumplost} appearing in 1.3K comments.
On closer examination, this hashtag seems to be used almost exclusively by left-leaning (and often clearly left-wing) accounts to troll Trump supporters.
  Finally, we see \texttt{\#transrights} and \texttt{\#transrightsarehumanrights} showing up in \numhttransrightsarehumanrights comments.
From manual examination, these again appear to primarily be engaged in trolling behavior.
  For example, a Ben Shapiro related parody account made use of \texttt{\#transrightsarehumanrights} (along with many other hashtags) as did an account posting various K-Pop related posts.
\jbnote{the above needs to be checked for phrasing and lies :3}
\jbnote{talk about ccp?}

\begin{table}[t]
\centering
\small
\begin{tabular}{lr}
\hline
\textbf{Hashtag}      & \textbf{\#Posts} \\ \midrule
covid         & 46,812   \\
trumprally    & 42,707   \\
trump         & 32,884   \\
gettr         & 27,950   \\
bitcoin       & 14,721   \\
trumpwon      & 11,581   \\
maga          & 11,516   \\
russia        & 10,454   \\
didi          & 10,054   \\
trump2024     & 8,953    \\
mudslide      & 7,677    \\
freedom       & 7,224    \\
america       & 7,160    \\
heatwave      & 6,463    \\
usa           & 6,270    \\
covid19       & 6,203    \\
biden         & 5,503    \\
bolsonaro2022 & 5,185    \\
news          & 4,776    \\ \bottomrule
\end{tabular}
\caption{Top 20 hashtags on user posts.}
\label{table:postshashtags}
\end{table}

\begin{table}[t]
  \centering
  \small
\begin{tabular}{lr}
\hline
\textbf{Hashtag}                  & \textbf{\#Comments} \\ \midrule
gettr                     & 7,465      \\ 
freedom                   & 4,269      \\ 
trumpwon                  & 4,173      \\ 
america                   & 4,063      \\ 
twitter                   & 3,642      \\ 
news                      & 3,536      \\ 
trump                     & 3,384      \\ 
trump2024                 & 2,019      \\ 
bolsonaro2022             & 1,542      \\ 
trumplost                 & 1,375      \\ 
trumprally                & 1,330      \\ 
ccp                       & 1,215      \\ 
saveamerica               & 959        \\ 
transrights               & 957        \\ 
covid                     & 891        \\ 
biden                     & 865        \\ 
wwg1wga                   & 844        \\ 
usa                       & 780        \\ 
transrightsarehumanrights & 762        \\ \bottomrule
\end{tabular}
\caption{Top 20 hashtags on user comments.}
\label{table:commentshashtags}
\end{table}

\descr{URL analysis.} %
Another mechanism for getting a high level view of an online community is the type of off-platform content its users link to.
For example, looking at the political leaning or trustworthiness of the most popular news outlets shared within a community can be used in a variety of higher order analyses~\cite{wang2021multi,zannettou2017web}.

In Table~\ref{table:postsurl} we show the the top 20 domains linked to in Gettr posts.
The most popular domain is \texttt{youtube.com}, and by a very wide margin (\posturlyoutube).
Prior to their deplatforming, 6.89\% of links on Gab~\cite{zannettou2018gab} and 13.59\% of links on Parler~\cite{aliapoulios2021large} pointed to YouTube\jbnote{can we compare to 4chan or something else? idk. we should also cite more papers than just ours. there are other papers that have done shit after us i think? can we cite dissenter paper here?}.
In addition to YouTube, links to alternative video sharing websites like Rumble (\urlrumble), bitchute (\urlbitchute)~\cite{trujillo2020bitchute}, and odysee (\urlodysee) are also quite popular\jbnote{we need cites here. at minimum ben horn has a bitchute paper, and probably some words about how these cites are part of the right-wing psycho committee}.

The only news sites appearing in the top 20 are conservative ones (Fox News) and well known peddlers of conspiracy theories, and mis- and dis-information (Epoch Times).
Gettr is notable here because of exclusively right-leaning news outlets.
For example, on Gab's Dissenter platform, the BBC was more popular than Fox News, and The Guardian also appeared in the top 10 most linked sites~\cite{rye2020reading}.

\jbnote{vk being there is perhaps interesting, idk. also frankspeech is that mypillow dipshits thing. also, we need to split posts and comments here into two separate tabels too please.}

Table~\ref{table:commentsurl} lists the top 20 domains that appeared in comments within our dataset.
We observe that Twitter links happen to be the second most shared links on the posts (\posturltwitter) and fourth most shared links on the comments (\commenturltwitter).
This is an interesting observation as one of the Gettr's primary goals is to become an alternative to Twitter.
We can also see sharing of external links to other social media such as Tiktok and Facebook.
We also find instances of far-right websites being pushed in the comments, such as \texttt{maga-patriot2024.com}, \texttt{rawconservativeopinions.com}, and \texttt{trumpbookusa.com}.

\begin{table}[t]
  \centering
  \small
\begin{tabular}{lr}
\hline
\textbf{Domain}           & \textbf{\#Posts} \\ \midrule
youtube.com          & 93,546   \\ 
twitter.com          & 35,091   \\ 
rumble.com          & 33,542   \\ 
thegatewaypundit.com & 16,030   \\ 
facebook.com         & 12,067   \\ 
instagram.com        & 10,201   \\ 
foxnews.com          & 9,755    \\ 
t.me             & 9,494    \\ 
theepochtimes.com    & 9,739    \\ 
bitchute.com         & 7,438    \\ 
gnews.org            & 6,899    \\ 
breitbart.com        & 6,104    \\ 
tiktok.com           & 5,150    \\ 
odysee.com           & 4,127    \\ 
vk.com               & 3,685    \\ 
gettr.com            & 3,530    \\ 
dailymail.co.uk       & 2,953    \\ 
newsmax.com          & 2,781    \\ 
nypost.com           & 2,681    \\ \bottomrule
\end{tabular}
\caption{Top 20 URL domains on user posts.}
\label{table:postsurl}
\end{table}

\begin{table}[t]
  \centering
  \small
\begin{tabular}{lr}
\hline
\textbf{Domain}                  & \textbf{\#Comments} \\ \midrule
youtube.com                 & 9,387       \\ 
gettr.com                   & 4,182       \\ 
rumble.com                  & 3,866       \\ 
twitter.com                 & 2,704       \\ 
theweeklyfront.com          & 1,788       \\ 
bitchute.com                & 1,589       \\ 
t.me                    & 1,360       \\ 
blogspot.com                & 1,102       \\ 
thegatewaypundit.com        & 759         \\ 
redpapernews.com            & 620         \\ 
facebook.com                & 583         \\ 
trumpbookusa.com            & 568         \\
gab.com                     & 508         \\
frankspeech.com             & 483         \\
thestrongestpost.com        & 415         \\
google.com                  & 413         \\
odysee.com                  & 401         \\
rawconservativeopinions.com & 397         \\
maga-patriot2024.com        & 337         \\ \bottomrule
\end{tabular}
\caption{Top 20 URL domains on user comments.}
\label{table:commentsurl}
\end{table}

\begin{figure*}[t]
\subfigure[Severe Toxicity]{\includegraphics[width=0.33\textwidth]{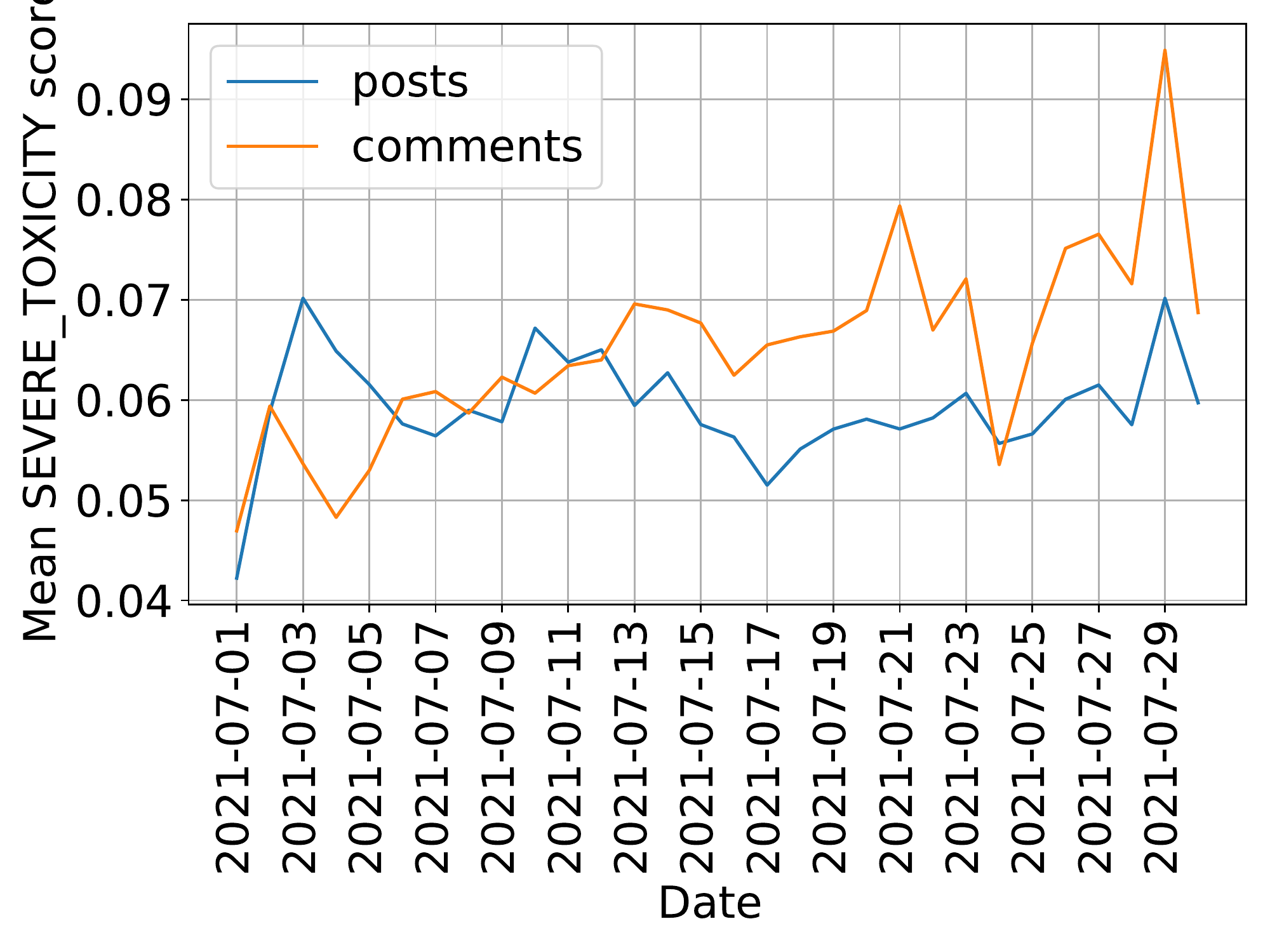}\label{fig:toxicity}}
\subfigure[Spam]{\includegraphics[width=0.33\textwidth]{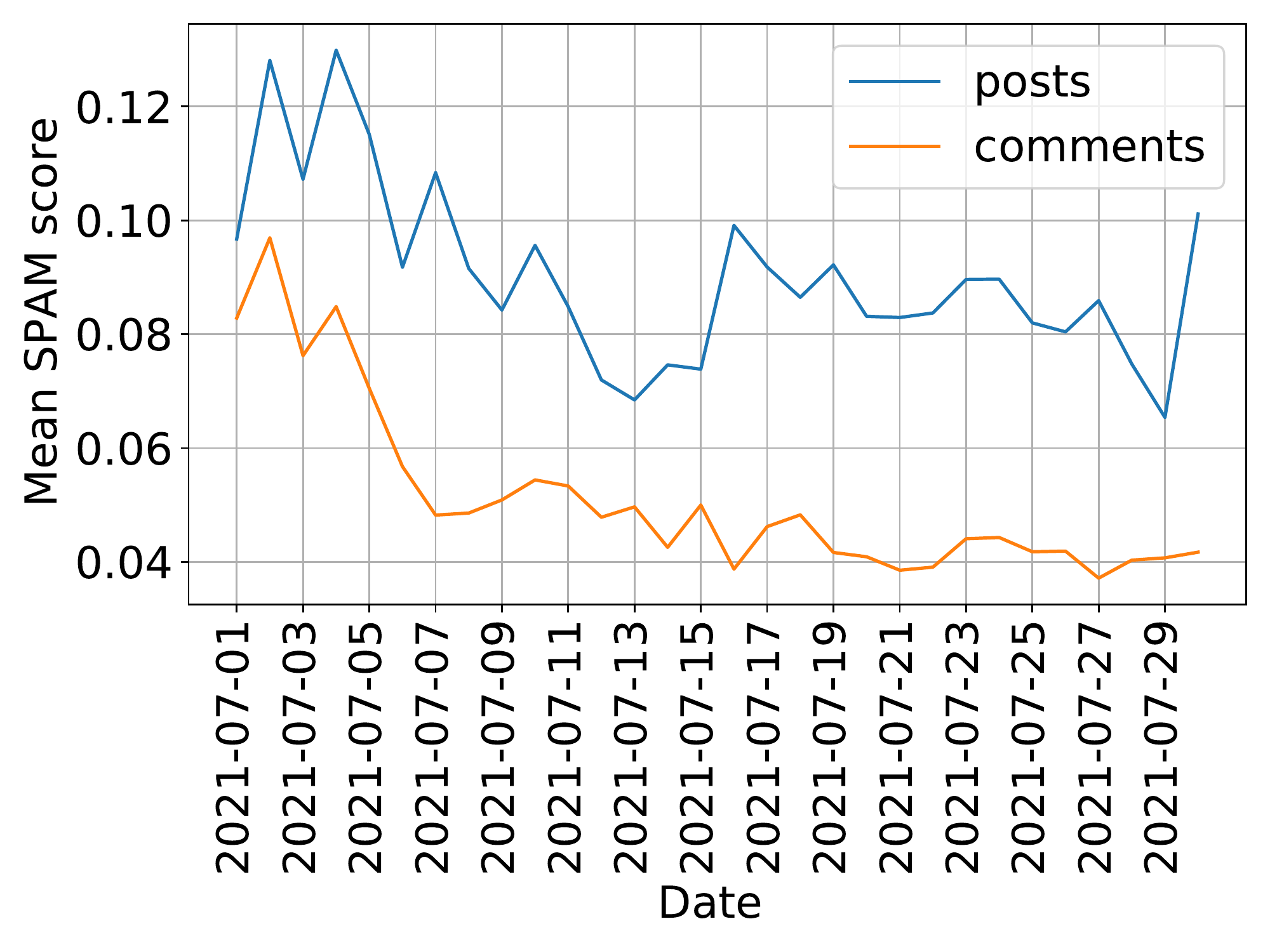}\label{fig:spam}}
\subfigure[Obscene]{\includegraphics[width=0.33\textwidth]{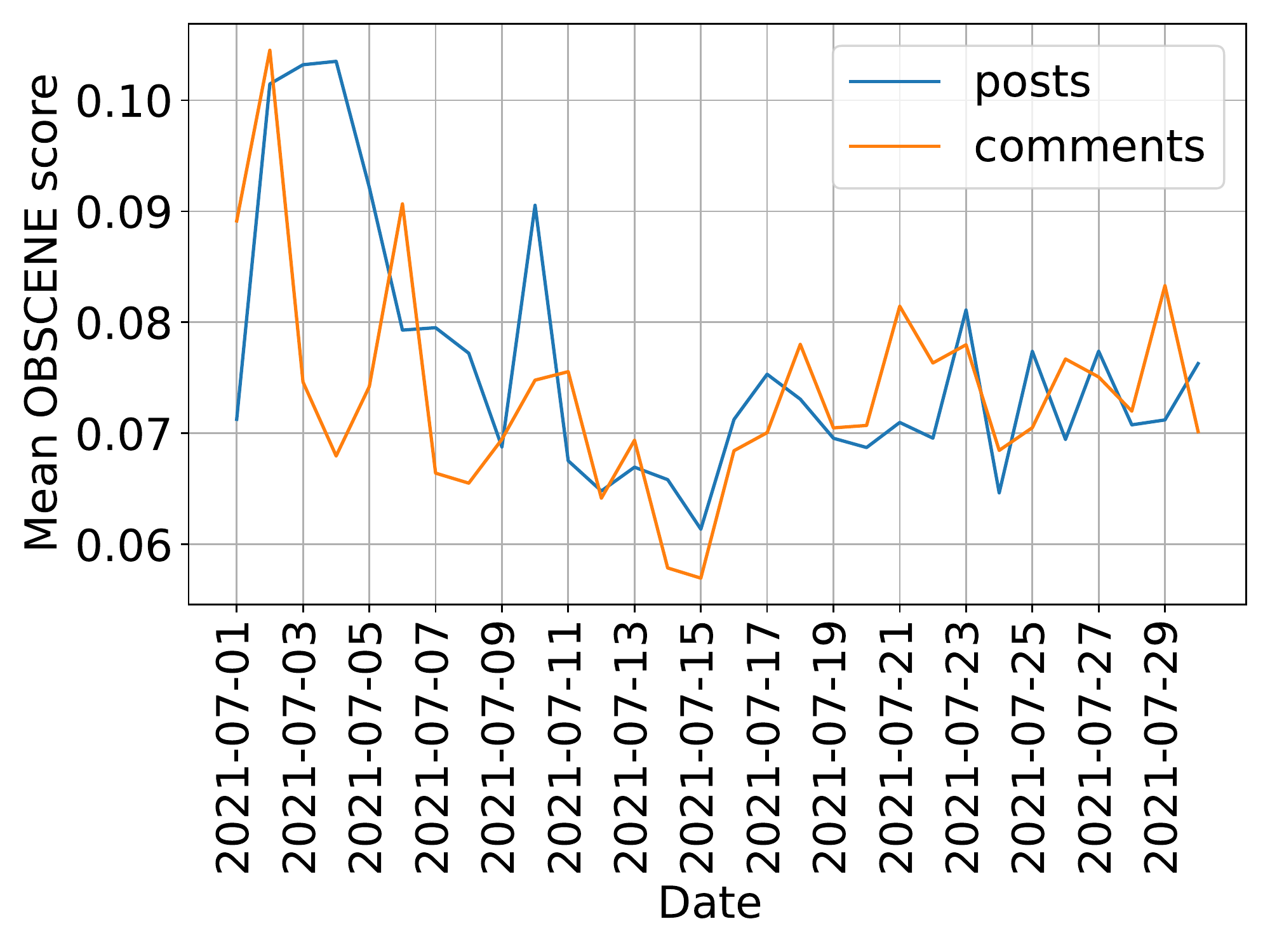}\label{fig:obscene}}
\subfigure[Identity Attack]{\includegraphics[width=0.33\textwidth]{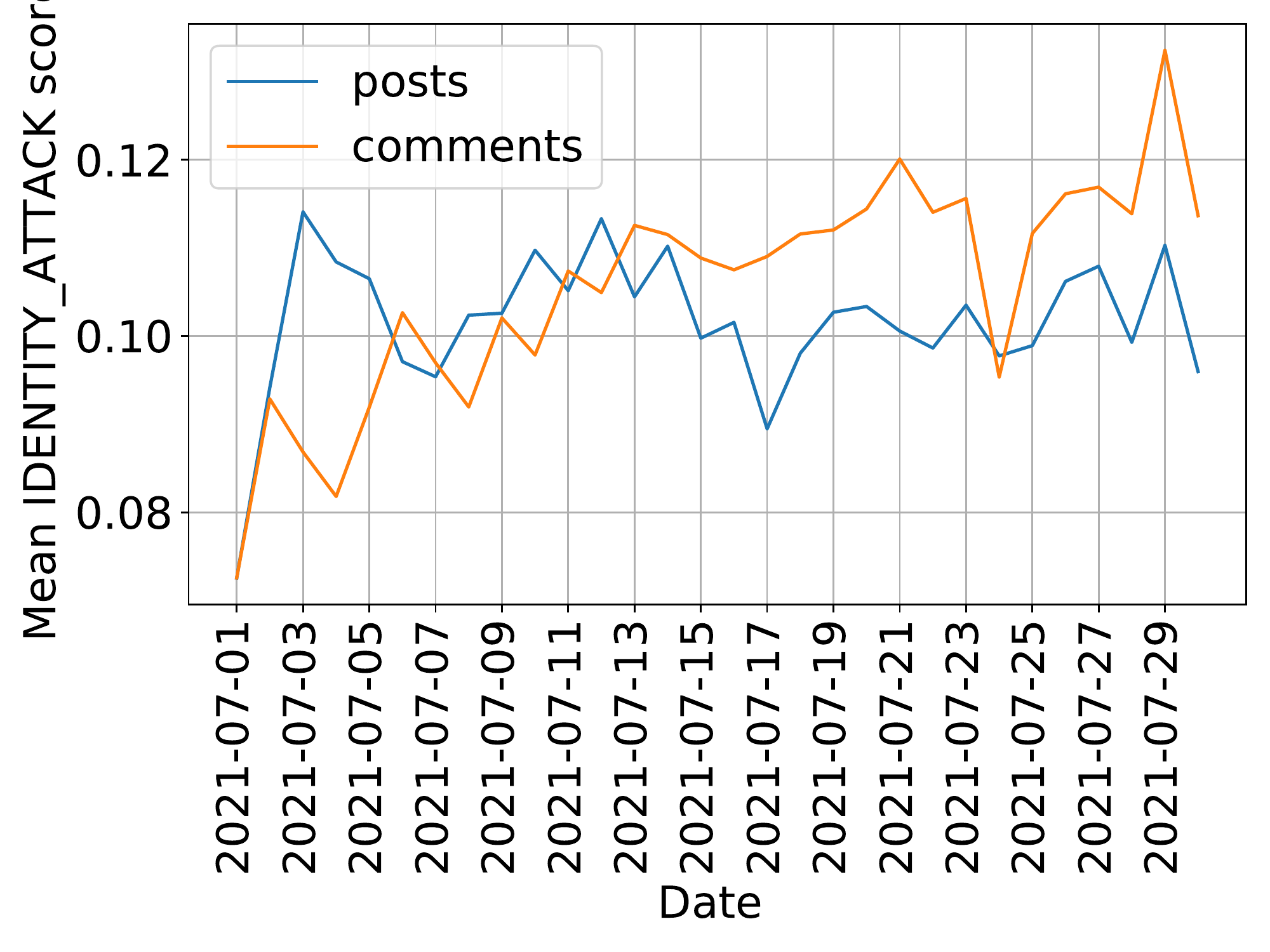}\label{fig:identity}}
\subfigure[Threat]{\includegraphics[width=0.33\textwidth]{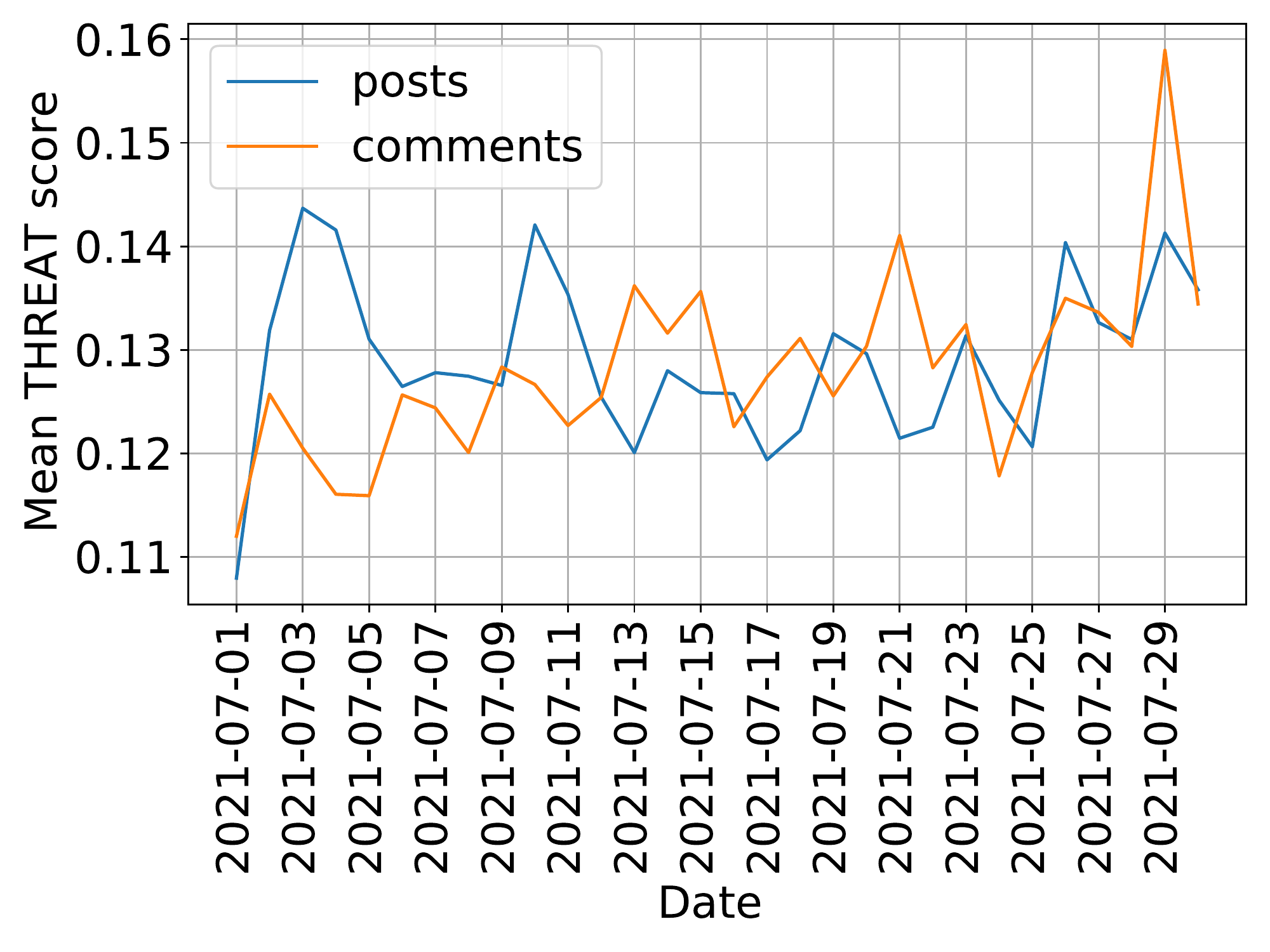}\label{fig:threat}}
\subfigure[Inflammatory]{\includegraphics[width=0.33\textwidth]{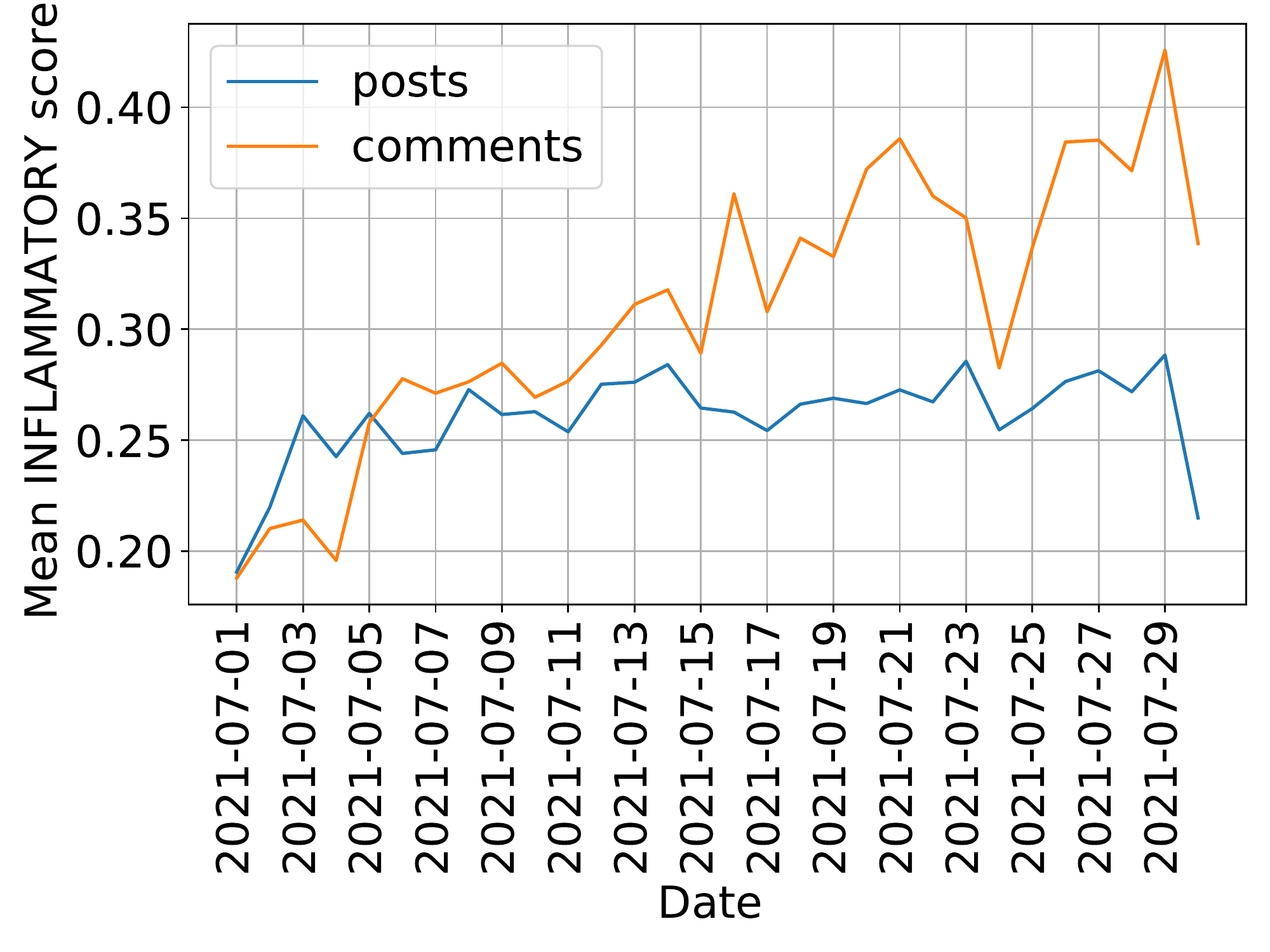}\label{fig:inflammatory}}
\caption{Average Perspective scores over time.}
\label{fig:perspective}
\end{figure*}

\descr{Progression of content.}
We use Google's Perspective API~\cite{perspectiveapi} to measure Gettr posts and comments in terms of toxicity, spam, profanity in the posts and comments.
We use the six different models  made available in the Perspective API:  1)~\textit{Severe toxicity} (Figure~\ref{fig:toxicity}) 2)~\textit{Spam} (Figure~\ref{fig:spam}) 3)~\textit{Identity Attack} (Figure~\ref{fig:identity}) 4)~\textit{Threat} (Figure~\ref{fig:threat}) 5)~\textit{Obscene} (Figure~\ref{fig:obscene}) 6)~\textit{Inflammatory} (Figure~\ref{fig:inflammatory}). %
We sampled 5,000 daily posts/comments and plot the daily mean of the Perspective scores in the figures. %

From Figure~\ref{fig:toxicity}, we see that, generally speaking, comments are more toxic than posts, and more specifically, toxicity in comments has a generally increasing trend.
That said, the average level of toxicity on Gettr is lower than what was recently reported by research on other alternative platforms like Gab~\cite{ali2021understanding} and 4chan~\cite{aliapoulios2021gospel}.
Comments and posts become less spammy over our observation period (Figure~\ref{fig:spam}), which might indicate that the platform has rolled out better anti-abuse systems, or alternatively, that spammers find little value in the platform.
The first possibility is potentially corroborated by the fact that the level of obscenity in posts and comments has also decreased since the platform went live, as seen in Figure~\ref{fig:obscene}.
In Figure~\ref{fig:identity} we see that there was an increase in the average score for identity attacks, with comments generally scoring higher, while in~\ref{fig:threat} this trend is not as clear.
Finally, we see that comments became increasingly inflammatory as seen in Figure~\ref{fig:inflammatory} while posts remained relatively stable.

\section{Conclusion}

In this paper, we took a first data-driven look at the Gettr social network.
We find that the user base of Gettr is similar to that of other alternative social network platforms like Gab and Parler.
We also find that while user activity has been steadily decreasing during the month of July 2021, there is a core of early adopters and verified users who keep being active on Gettr. 
Our analysis identified several interesting directions that could be explored as future work.
First, the crystallization of the activity on Gettr around a core of verified users and early adopters could give interesting insights into how like-minded online communities form and evolve.
Second, it would be interesting to analyze the presence of disinformation and conspiratorial content in more depth, to better understand its effect on online discourse.

\small
\bibliographystyle{abbrv}
\bibliography{bibilography.bib}

\end{document}